\documentclass[twocolumn]{aastex62}

\newcommand{\lya}{\mathrm{Ly}\ensuremath{\alpha}}
\newcommand{\nv}{\ion{N}{5}}

\newcommand{\cii}{[\ion{C}{2}]}
\newcommand{\ciib}{\ion{C}{2}}
\newcommand{\mgii}{\ion{Mg}{2}}
\newcommand{\civ}{\ion{C}{4}}
\newcommand{\siiv}{\ion{Si}{4}}
\newcommand{\siii}{\ion{Si}{2}}
\newcommand{\hi}{\ion{H}{1}}

\newcommand{\alii}{\ion{Al}{2}}
\newcommand{\ciii}{\ion{C}{3}]}
\newcommand{\feii}{\ion{Fe}{2}}
\newcommand{\oi}{\ion{O}{1}}
\newcommand{\ovi}{\ion{O}{6}}

\newcommand{\fhi}{\overline{x}_{\mathrm{H}\ensuremath{\textsc{i}}}}

\newcommand{\kms}{{\rm km\,s}\ensuremath{^{-1}}}

\accepted{in ApJ, September 2, 2019}

\shorttitle{A metal-poor damped $\lya$ system at redshift 6.4}
\shortauthors{Ba\~nados et al.}

\begin{document}

\title{A metal--poor damped $\lya$ system at redshift 6.4}

\correspondingauthor{Eduardo Ba\~nados}
\email{banados@mpia.de}

\author[0000-0002-2931-7824]{Eduardo Ba\~nados}
\affiliation{The Observatories of the Carnegie Institution for Science, 813 Santa Barbara Street, Pasadena, CA 91101, USA}
\affiliation{Max-Planck-Institut f\"{u}r Astronomie, K\"{o}nigstuhl 17, D-69117, Heidelberg, Germany}

\author{Michael Rauch}
\affiliation{The Observatories of the Carnegie Institution for Science, 813 Santa Barbara Street, Pasadena, CA 91101, USA}

\author[0000-0002-2662-8803]{Roberto Decarli}
\affiliation{INAF -- Osservatorio di Astrofisica e Scienza dello Spazio, via Gobetti 93/3, I-40129, Bologna, Italy}

\author[0000-0002-6822-2254]{Emanuele~P.~Farina}
\affiliation{Department of Physics, Broida Hall, University of California, Santa Barbara, CA 93106--9530, USA}
\affiliation{Max-Planck-Institut f\"{u}r Astronomie, K\"{o}nigstuhl 17, D-69117, Heidelberg, Germany}

\author[0000-0002-7054-4332]{Joseph~F.~Hennawi}
\affiliation{Department of Physics, Broida Hall, University of California, Santa Barbara, CA 93106--9530, USA}
\affiliation{Max-Planck-Institut f\"{u}r Astronomie, K\"{o}nigstuhl 17, D-69117, Heidelberg, Germany}

\author[0000-0002-5941-5214]{Chiara Mazzucchelli}

\affiliation{European Southern Observatory, Alonso de Cordova 3107, Vitacura, Region Metropolitana, Chile}

\author[0000-0001-9024-8322]{Bram P. Venemans}
\affiliation{Max-Planck-Institut f\"{u}r Astronomie, K\"{o}nigstuhl 17, D-69117, Heidelberg, Germany}

\author[0000-0003-4793-7880]{Fabian Walter}
\affiliation{Max-Planck-Institut f\"{u}r Astronomie, K\"{o}nigstuhl 17, D-69117, Heidelberg, Germany}
\affiliation{National Radio Astronomy Observatory, Pete V. Domenici Array Science Center, P.O. Box 0, Socorro, NM 87801, USA}

\author[0000-0003-3769-9559]{Robert A. Simcoe}
\affiliation{MIT-Kavli Center for Astrophysics and Space Research, 77 Massachusetts Avenue, Cambridge, MA, 02139, USA}

\author[0000-0002-7738-6875]{J. Xavier Prochaska}
\affiliation{Department of Astronomy and Astrophysics, University of California, Santa Cruz, CA 95064, USA}

\author{Thomas Cooper}
\affiliation{The Observatories of the Carnegie Institution for Science, 813 Santa Barbara Street, Pasadena, CA 91101, USA}

\author[0000-0003-0821-3644]{Frederick B. Davies}
\affiliation{Department of Physics, Broida Hall, University of California, Santa Barbara, CA 93106--9530, USA}

\author{Shi-Fan~S.~Chen}
\affiliation{MIT-Kavli Center for Astrophysics and Space Research, 77 Massachusetts Avenue, Cambridge, MA, 02139, USA}
\affiliation{Department of Physics, University of California, Berkeley, CA 94720, USA}

\begin{abstract}

We identify a strong Ly$\alpha$ damping wing profile in the spectrum of the quasar P183+05 at $z=6.4386$. Given the detection of several narrow metal absorption lines at $z=6.40392$, the most likely explanation for the absorption profile is that it is due to a damped $\lya$ system. However, in order to match the data a contribution of an intergalactic medium $5-38\%$ neutral or additional weaker absorbers near the quasar is also required. The absorption system  presented here is the most distant damped $\lya$ system currently known. We estimate an \hi\ column density of $10^{20.68\pm0.25}\,$cm$^{-2}$, metallicity [O/H]$=-2.92\pm 0.32$, and relative chemical abundances of a system consistent with a low-mass galaxy during the first Gyr of the universe. This object is among the most metal-poor damped $\lya$ systems known and, even though it is observed only $\sim$$850\,$Myr after the big bang, its relative abundances do not show signatures of chemical enrichment by Population III stars.
\end{abstract}




\section{Introduction} \label{sec:intro}

The epoch of reionization started a few hundred million years after the big bang \citep[e.g.,][]{greig2017a}   when the collapse of the first dark matter halos and gas cooling led to the formation of the first generation of stars (Population III stars) and galaxies \citep{dayal2018}. These sources are thought to play an important role in the production of the high-energy photons required to reionize the intergalactic medium (IGM) and end the cosmic dark ages within the first billion years of the universe \citep[e.g.,][]{fan2006c}.

Nevertheless, our understanding of the properties of the Population III stars is limited and mainly based on theoretical models \citep{glover2013,greif2015}, while their direct observational characterization is likely beyond the capabilities of existing telescopes. Current observational efforts focus on the study of chemical abundances of very metal-poor stars, dwarf galaxies, and high-redshift gas clouds, which could still retain the chemical enrichment signatures produced by the first and second generation of stars \citep{frebel2015,hartwig2018,jeon2019}. The high-redshift clouds are observed as damped $\lya$ absorber (DLA) systems with high \hi\ column density ($N_{\mathrm{HI}} > 2\times 10^{20}\,$cm$^{-2}$) along the line of sight of background high-redshift quasars and gamma ray bursts (GRBs).

DLAs are thought to be associated with low-mass galaxies (or protogalaxies) at the faint end of the luminosity function \citep{haehnelt2000}. Importantly, metal-poor DLAs have been suggested to be the progenitors of present-day dwarf galaxies \citep{cooke2015}. The most distant of such systems ($z>6$) may thus hold clues for constraining the initial mass function of Population III stars and their contribution to reionization \citep{kulkarni2013,kulkarni2014,ma2017b}.
A complication is that at $z>5$ the high opacity of the $\lya$ forest makes it nearly impossible to measure the \hi\ column density of high-redshift absorption systems \cite[e.g.,][]{beckerG2012,rafelski2014} unless they are located in the ``proximity zone'' of the quasar (i.e., within $\sim$$5000\,\kms$) or they correspond to GRB host galaxies.
We note that these ``proximate DLAs'' (PDLAs) are  often excluded from analyses of DLAs in case their properties could be  affected by the quasar radiation. However, it has been argued that PDLAs are probably \textit{not} associated with the quasar hosts.
This is based on the significant quasar--absorber separations inferred from studies of the \siii$^*$ and \ciib$^*$ fine-structure lines  \citep{ellison2010}.

 PDLAs in quasar spectra are already quite rare at $z\sim 3$ \citep{prochaska2008} but with the increasing number of quasars discovered at $z\gtrsim 6$ \cite[e.g.,][]{banados2016,yang2019b} it is not unexpected to find the first PDLA examples at such high redshifts. Until recently,  the only DLA-like absorption profiles observed in quasars at $z\gtrsim 5.2$ were in the two most distant quasars currently known at $z>7$ \citep{mortlock2011,banados2018a}. However, for these two cases no metal lines associated with a potential DLA are detected \citep{simcoe2012,banados2018a} and the most likely explanation is that the measured absorption profile is caused by a significantly neutral ($>10\%$) IGM \citep{miralda-escude1998}. Indeed, the IGM damping wings observed in these quasars provide some of the strongest constraints on the average hydrogen neutral fraction ($\fhi$) in the epoch of reionization (\citealt{bolton2011,greig2017a,davies2018b}; but see also \citealt{greig2019}).
The first detection of a PDLA along the line of sight of a $z\gtrsim 5.2$ quasar was recently reported by \cite{dodorico2018} toward the $z=6.0025$ quasar SDSS~J2310+1855.

In this paper we report a system similar to that identified by \cite{dodorico2018}: a metal-poor DLA at $z=6.40392$ along the line of sight to the $z=6.4386$ quasar PSO~J183.1124+05.0926 (hereafter P183+05; R.A.$=12^{\rm h} 12^{\rm m} 26\fs981$; decl.$=+05^{\circ} 05^{\prime} 33 \farcs 49$).
This paper is structured as follows.
In Section \ref{sec:observations} we introduce the data used and the detection of the $\lya$ damping wing. In Section \ref{sec:dla-metals} we argue that the most plausible explanation for this damping wing is that it is  produced by a DLA in combination of a surrounding neutral IGM or additional weaker absorbers.  In Section \ref{sec:relabund} we measure the relative abundances for the DLA, while in Section \ref{sec:cont-blue} we estimate its metallicity. We discuss our results in  Section \ref{sec:discussion}, including the low metallicity of this DLA and the possible scenarios that yielded its observed chemical patterns.
Finally, we summarize our results and conclusions in Section \ref{sec:summary}.

We use a flat $\Lambda$CDM cosmology
with $H_0 = 67.7 \,\mbox{km\,s}^{-1}$ Mpc$^{-1}$, $\Omega_M = 0.307$, and $\Omega_\Lambda = 0.693$ \citep[][]{planck2016xiii}. In this cosmology, the universe was 857 Myr old at $z=6.4$.

\begin{figure*}[!ht]
\centering
\figurenum{1}
\plotone{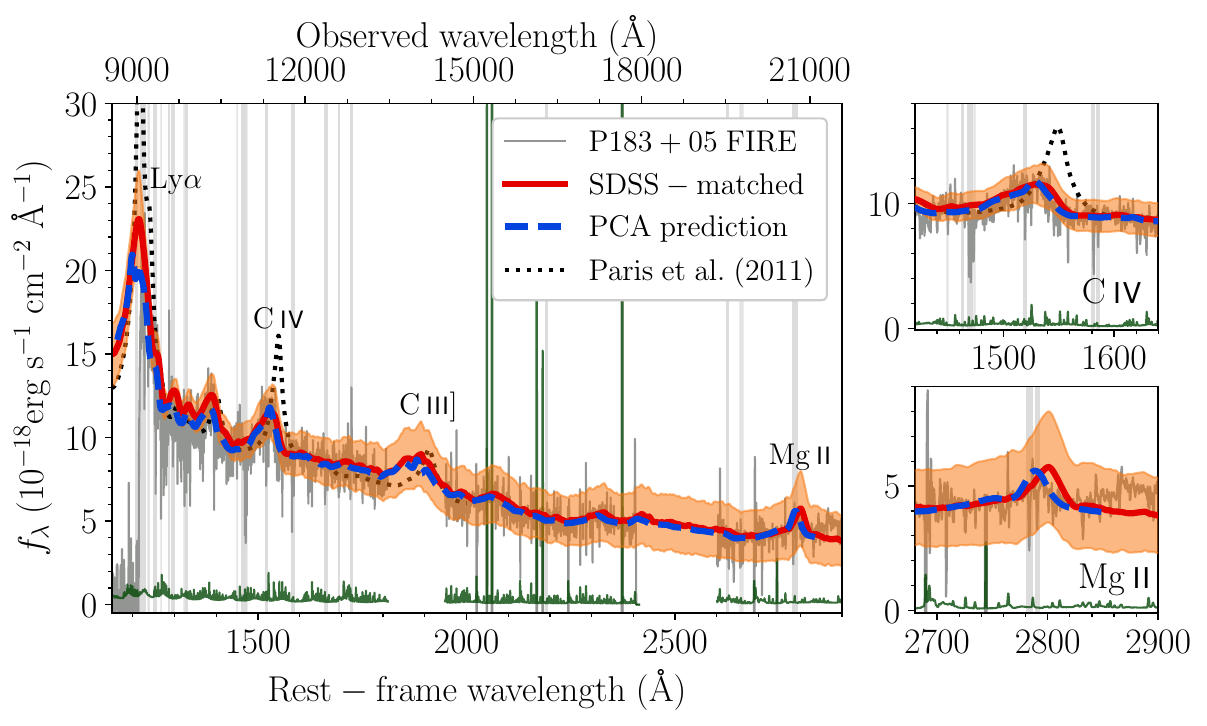}
\caption{
FIRE spectrum of P183+05 (gray line) and its $1\sigma$ error vector (green line).  The red line is the mean spectrum of low-redshift SDSS quasar with \civ\ properties matched to P183+05 (see Section \ref{sec:cont-blue} for details) and the orange region represents the $\pm1\sigma$ dispersion around the mean among the 61 spectra of the P183+05 analogs used to create the composite spectrum. The blue dashed line is the principal component analysis (PCA) predicted continuum using the method of \cite{davies2018a}. The vertical gray-shaded regions were masked out when creating the SDSS-matched and PCA spectra to avoid foreground absorbers and strong sky-subtraction residuals.
The foreground absorbers are at $z=6.40392$, $z=6.0645$, $z=5.8434$, $z=5.0172$, $z=3.4185$, and  $z=3.2070$. In this paper we focus on the highest-redshift absorber (determined from metal absorption features, see Section \ref{sec:dla-metals}).
Both spectra match the general properties observed in the spectrum of P183+05 redward of the $\lya$ emission line. The dotted line is the mean SDSS quasar spectrum from \cite{paris2011}, which shows stronger emission lines than P183+05. The top and bottom panels on the right-hand side show a zoom-in to the regions of the \civ\ and \mgii\ lines, respectively.
}

\label{fig:sdssmatch}
\end{figure*}

\begin{figure*}[!ht]
\centering
\figurenum{2}
\plotone{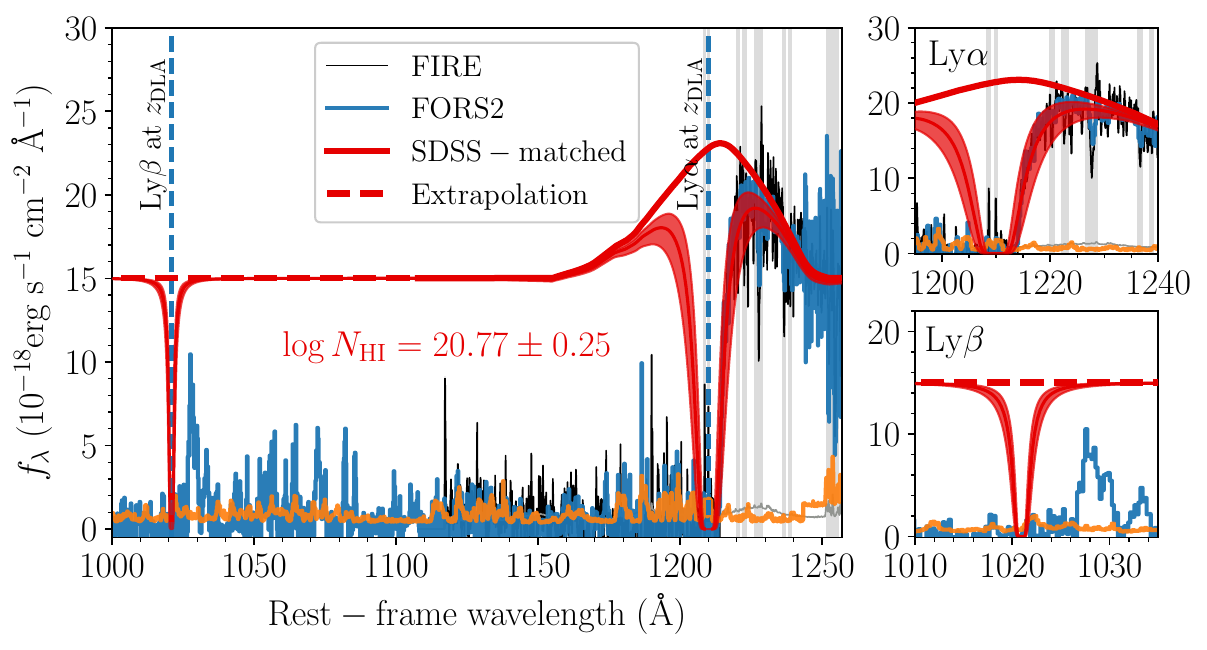}
\caption{
FORS2 spectrum of P183+05 (blue line) and  its $1\sigma$ error vector (orange line).
The $x$-axes show the rest-frame wavelength at the redshift of the quasar, $z=6.4386$.
The FIRE spectrum and masked regions from Figure \ref{fig:sdssmatch} are shown for comparison.
 The vertical dashed lines show the expected position of the $\lya$ and Ly$\beta$ lines at the redshift of the DLA,  $z_{\rm DLA}=6.40392$ (see Section \ref{sec:dla-metals}). The thick solid red line is the mean spectrum of SDSS-matched quasars (see also Figure \ref{fig:sdssmatch}) and the dashed red line is a simple extrapolation to cover the Ly$\beta$ region.   The red shaded regions show absorption caused by a $z=6.40392$ DLA with a column density of hydrogen of $N_{\mathrm{HI}}=10^{20.77\pm 0.25}\,$cm$^{-2}$.  The top and bottom panels on the right-hand side show a zoom-in to the regions of $\lya$ and Ly$\beta$, respectively. The absorption profile seen in the FORS2 data is consistent with the FIRE data and it provides additional information in the Ly$\beta$ region.
}
\label{fig:fors2}
 \end{figure*}

 \begin{figure*}[!ht]
\centering
\figurenum{3}
\plotone{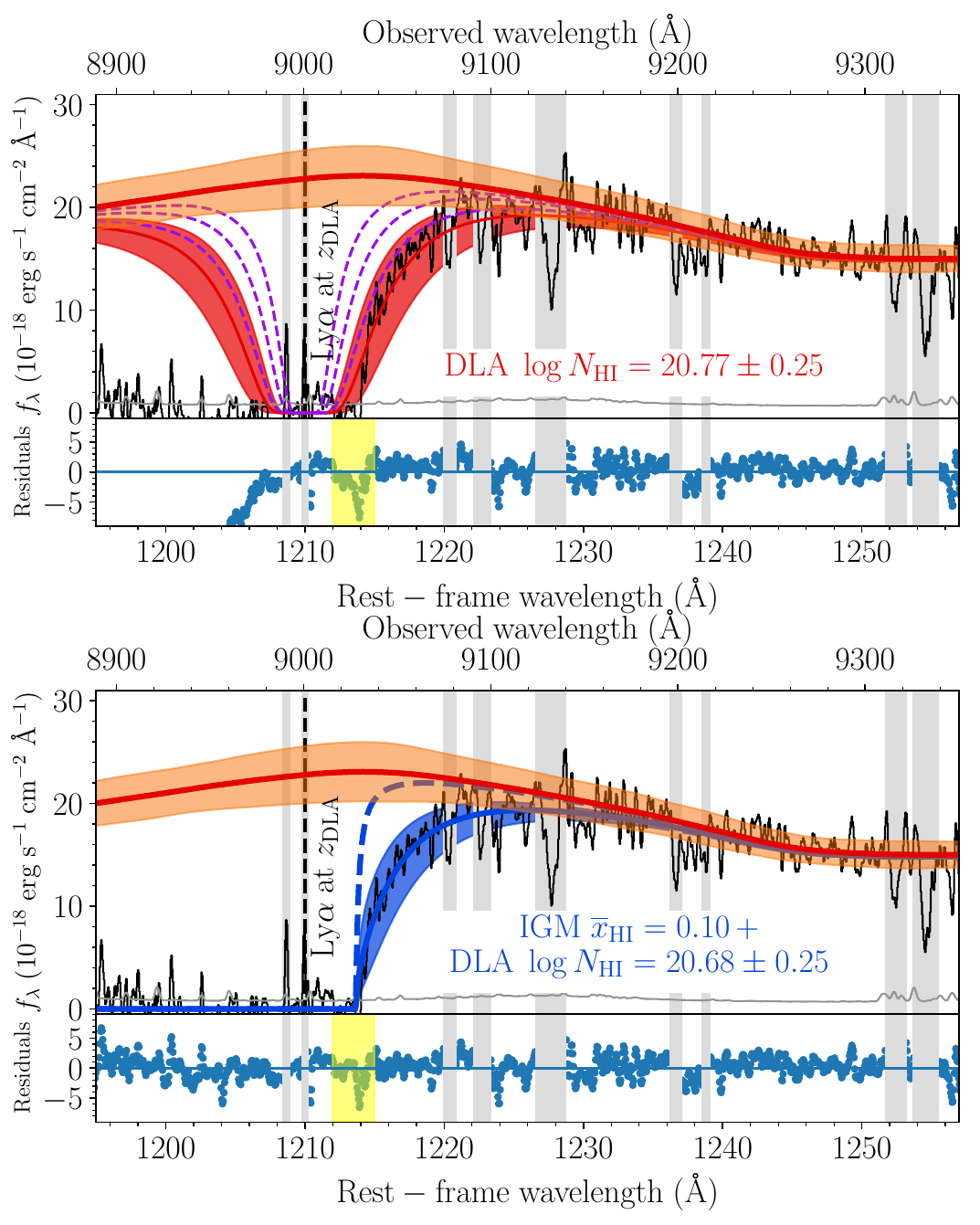}
\caption{
Both panels: FIRE spectrum of P183+05 (black line) and its $1\sigma$ error vector (gray line).
The vertical gray-shaded regions were masked out to avoid foreground absorbers and sky-subtraction residuals. The $x$-axes show the rest-frame wavelength at the redshift of the quasar, $z=6.4386$.
The red line is the mean SDSS-matched quasar spectrum and the orange region represents the $\pm1\sigma$ dispersion among P183+05-analogs used to create the composite spectrum. The vertical dashed lines show the expected position of the $\lya$ line at the redshift of the DLA  $z_{\rm DLA}=6.40392$ (see Section \ref{sec:dla-metals}). Top: the red line shows a DLA model with $N_{\mathrm{HI}}=10^{20.77\pm 0.25}\,$cm$^{-2}$. Even though it seems a reasonable fit, it cannot reproduce the sharp drop in flux around $\lambda_{\rm rest}=1212-1215\,$\AA\ (see yellow region in the residuals panel). {The violet dashed lines correspond to DLA profiles with $\log N_{\mathrm{HI}}=$20.0, 20.3, and 20.6  for visual aid on the effect of  varying $N_{\mathrm{HI}}$}.
  Bottom: the dashed blue line represents the attenuation caused by an IGM with $\fhi=0.10$, while the solid blue line shows the best-fit DLA model using as input the dashed blue line, yielding $N_{\mathrm{HI}}=10^{20.68\pm 0.25}\,$cm$^{-2}$. A joint IGM+DLA fit better reproduces the data around $\lambda_{\rm rest}=1212-1215\,$\AA\ (see also Appendix \ref{app:igmpdla}).
}
\label{fig:dlaigm}
\end{figure*}

\section{Observations and $\lya$ damping wing} \label{sec:observations}

The quasar P183+05
was selected as a $z$-dropout in the Pan-STARRS1 survey \citep{chambers2017}. The details of its discovery and properties are presented in \cite{mazzucchelli2017b} and here we just summarize its main characteristics. P183+05 is a luminous quasar at $z=6.4$ with an AB $J$-band magnitude of $19.77\pm0.08$, and rest-frame 1450\,\AA\ apparent and absolute magnitudes of $19.82$ and $-27.03$, respectively. In this paper we use the spectrum taken with the Folded-port InfraRed Echellette \citep[FIRE;][]{simcoe2013} at the Baade telescope in Las Campanas Observatory on 2015 April 6. The quasar was observed for $11730\,$s in the echellette mode with the $0\farcs6$ slit, yielding a spectral resolution of $R=6000$ ($\sim$$50\,\kms$) over the range $8000-23000\,$\AA.
In order to study the quasar's Ly$\beta$ region we used the spectrum taken with the Focal Reducer Low-Dispersion Spectrograph 2 \citep[FORS2;][]{appenzeller1998} at the Very Large Telescope. We used the GRIS-600z grism in combination with the OG590 filter. The slit width was $1\farcs3$ which resulted in a spectral resolution of $R\sim$1000. The wavelength range was $7100-10400$\,\AA. The pixels were binned 2x2, giving a spatial scale of $0\farcs25\,$pixel$^{-1}$ and a dispersion of 1.62\,\AA\,pixel$^{-1}$.
This spectrum was observed for $2550$\,s on 2015 May 8.
The
modest quality of the existing data prevents us from estimating an accurate \mgii-based redshift and black hole mass (see \citealt{mazzucchelli2017b}).
However, the \cii\ emission from the quasar host galaxy is well detected with ALMA, yielding an accurate systemic redshift of $z=6.4386 \pm 0.0004$ \citep{decarli2018}.

Figure \ref{fig:sdssmatch} shows the FIRE spectrum for P183+05 and Figure \ref{fig:fors2} shows both FIRE and FORS2 spectra in the Ly$\beta$ and $\lya$ regions. The spectrum shows a strong $\lya$ damping wing profile that resembles the absorption caused by a DLA with a column density of hydrogen $N_{\mathrm{HI}}\gtrsim10^{20.5}\,$cm$^{-2}$; such a strong column density is also compatible with the trough observed in the Ly$\beta$ region (see Figure \ref{fig:fors2}). However, without any further information it is not possible to distinguish between an absorption profile caused by a DLA or by an IGM damping wing.  In fact, without any priors, an absorption profile caused by a very neutral IGM ($\fhi=0.80$) fits the data better than a single DLA (see Appendix \ref{app:igmpos}).

In the following, we will argue that this particular absorption profile is more likely to be caused by a combination of a DLA and either a relatively neutral environment or additional weaker absorbers near the quasar. This system is similar to the absorber recently reported by \cite{dodorico2018} at $z=5.94$. Unlike the $z=5.94$ absorber, which was serendipitously detected in CO(6--5) emission by ALMA, there is no evidence for any other source besides the bright quasar host galaxy in the available ALMA observations of the P183+05 field \citep{decarli2018,champagne2018}.
Recent deeper ALMA observations of this field centered on the quasar's \cii\ $158\,\mu\mathrm{m}$ emission line detect two additional continuum sources but no significant \cii\ emission at the redshift of the DLA reported here (M.~Neeleman et al.\ in preparation).

\section{Proximate Damped $\lya$ System} \label{sec:dla-metals}

\begin{figure*}[!ht]
\centering
\figurenum{4}
\plotone{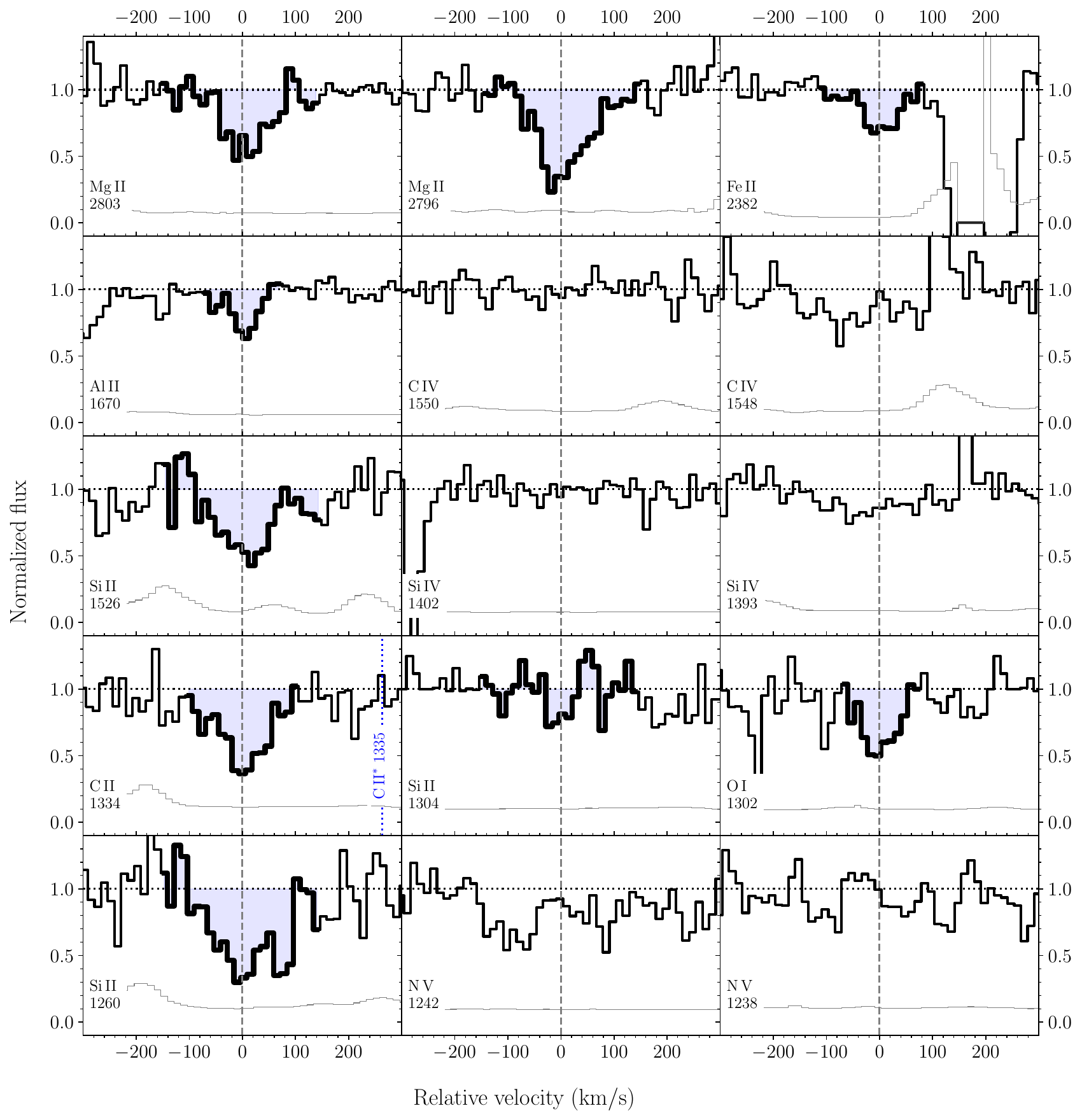}
\caption{
 Continuum-normalized spectral regions centered at a rest velocity corresponding to an absorber at $z=6.40392$ (dashed vertical lines). The blue shaded areas show the regions used to measure their column densities (see Table \ref{tab:abundances}).
 The high excitation lines \civ, \siiv, and \nv\ are not detected but their expected locations are shown here for completeness (see also Table \ref{tab:ews}).  The horizontal dotted lines mark the continuum level, and the gray lines represent the $1\sigma$ uncertainties. The dotted blue line in the panel of the \ciib\ $\lambda 1334$ line corresponds to the expected position of the \ciib$^*$ line, which is not detected in this data (see text for implications).
The  \siii\ $\lambda 1260$ transition is potentially contaminated by a \civ\ 1548 absorption line from a system at $z=5.0172$, while the strength of \siii\ $\lambda 1526$  overpredicts the observed strength of the marginal ($\sim$$3\sigma$) detection of \siii\ $\lambda 1304$ (see Appendix \ref{app:saturation}).
}
\label{fig:metals}
 \end{figure*}

The large wavelength coverage provided by the FIRE spectrum allows us to investigate whether there is an absorber near the quasar that could produce the $\lya$ damping wing observed in Figures \ref{fig:fors2} and  \ref{fig:dlaigm}. In fact, we identify a \mgii\ $\lambda \lambda 2796, 2803$ doublet at $\lambda=(20704.2\,\mbox{\AA}, 20757.1\,\mbox{\AA})$. This corresponds to an absorber at $z=6.40392 \pm 0.0005$, which is further confirmed by the presence of additional metal absorption lines of \feii, \alii, \siii, \ciib, and \oi\ at the same redshift (see Figure \ref{fig:metals}). This \mgii\ system was independently identified by \cite{chen2017},
 who reported a redshift of $z=6.404$ for the absorber, consistent with our measurement.
The redshift difference between the quasar and the DLA is $\Delta z=0.0347$, which corresponds to a mere $1398\,\kms$ or 1.8 physical Mpc if they are in the Hubble flow (not necessarily a valid assumption; see e.g., \citealt{ellison2010}).

Based on the presence of discrete narrow metal absorption lines we argue that absorption by a PDLA cloud is the most likely explanation for the $\lya$ damping wing observed in the spectrum of P183+05.
However, we will show in Section~\ref{sec:cont-blue}
that the data favor a scenario where the absorption profile is caused by the joint effects of a DLA and a surrounding IGM with $\fhi=0.05-0.38$.
These data enable us to measure the relative abundances of this DLA as well as constrain its neutral hydrogen column density, allowing us to perform the first direct measurement of metallicity in a galaxy at $z>6$.

\subsection{Column densities and relative abundances}
\label{sec:relabund}

To estimate the column densities of the metals shown in Figure \ref{fig:metals}, we first need to model the quasar continuum in order to normalize the FIRE spectrum. We use the interactive task \textit{continuumfit} from the \textit{linetools} python package. The continuum is  fit by interpolating cubic splines between knots placed along the spectrum.

We visually inspect the normalized spectrum for all potential metal absorption lines consistent with being at $z=6.43902$ and select the velocity limits that are used for subsequent analysis (see Figure \ref{fig:metals}). Next, we calculate the rest-frame equivalent widths (EWs) and column densities. All these quantities are listed in Table \ref{tab:ews}. We calculate the column densities with the apparent optical depth method \citep[AODM;][]{savage1991}, using the wavelengths and oscillator strengths reported in \cite{morton2003}.

At intermediate resolution ($R\sim 6000$), saturation or blending of lines can show up as discrepant column density estimates for different transitions of the same ion.
This is the case for the \siii\ detection in our system: the column densities for \siii\ $\lambda 1260$ and  \siii\ $\lambda 1526$ are $10^{13.53}$\,cm$^{-2}$ and $10^{14.15}$\,cm$^{-2}$, respectively (see Table \ref{tab:ews}). As discussed in Appendix \ref{app:saturation}, the \siii\ $\lambda 1526$ line must be contaminated as its column density predicts much stronger \siii\ $\lambda 1260$ and $\lambda 1304$ absorptions than observed (see Figure \ref{fig:app_si}).
We therefore do not use the \siii\ $\lambda 1526$ column density in the remainder of the paper.
The column density of \siii\ $\lambda 1260$ is consistent with the marginal detection of \siii\ $\lambda 1304$; however, the \siii\ $\lambda 1260$ column density is likely to be contaminated by a \civ\ $\lambda 1548$ absorption line from a system at $z=5.0172$. Therefore, we consider the \siii\ $\lambda 1260$ measurement as an upper limit.
In Appendix \ref{app:saturation},
we conclude that we cannot rule out the possibility of hidden saturation for \oi\ $\lambda 1302$ and \alii\ $\lambda 1670$. Thus, to take possible saturation effects into account, we have increased their uncertainties from 0.06 to 0.20\,dex (see Table \ref{tab:ews} and Figures \ref{fig:app_oi} and \ref{fig:app_al}). This highlights the need for obtaining much higher-resolution spectroscopy in systems like this one, although this is very challenging for the current generation of telescopes.

 We note that the fine-structure line \ciib$^*$ $\lambda 1335$ is not detected in our data (dotted blue line in Figure \ref{fig:metals}). This line is ubiquitous in the spectra of GRB host galaxies (GRB-DLAs; e.g.,  \citealt{fynbo2009}).
 Under some assumptions, \ciib$^*$ can be used to determine the distance between quasar and absorber (we refer the reader to the discussion in \citealt{ellison2010} and references therein). The non-detection of this line supports the idea that this PDLA is not associated with the quasar host galaxy.

In Figure \ref{fig:metals} we also show the regions around \civ\ $\lambda\lambda 1548, 1550$, \siiv\ $\lambda\lambda 1393,1402$, \nv\ $\lambda\lambda 1238,1242$. They are not convincingly detected even though the AODM in the velocity range ($-150\,\kms$, $150\,\kms$) reports that \civ\ $\lambda 1548$, \siiv\ $\lambda1393$, and  \nv\ $\lambda 1242$ are more than $3\sigma$ significant. The oscillator strength of \nv\ $\lambda 1242$ is weaker than that of \nv\ $\lambda 1238$, therefore the possible absorption at $\lambda 1242$ is inconsistent with the non-detection of $\lambda 1238$.   Higher signal-to-noise ratio (S/N) data are required to confirm potential detection of high-excitation lines but in the remainder of the paper we consider these lines as non-detections and report their $3\sigma$ limits (Table \ref{tab:ews}).
In Table \ref{tab:abundances} we show the element ratios relative to the solar abundances reported in  \cite{asplund2009}. Solar abundances are defined using the meteoritic values for all elements other than C and O, for which we use the photospheric values given that these elements are volatile and cannot be recovered completely from meteorites \citep{lodders2003}.

\subsection{$\lya$ modeling and metallicity}
\label{sec:cont-blue}
In order to estimate the \hi\ column density of the absorber at $z=6.40392$, we first need to model the quasar's intrinsic $\lya$ emission.
This is not straightforward given the variety of $\lya$ strengths observed among quasars.
A further complication for P183+05 is that its \civ\ line is blueshifted by $5057 \pm 93\, \kms$ with respect to the systemic \cii\ redshift.  Quasar spectra with such large blueshifts are not rare among the highest-redshift quasars \citep{mazzucchelli2017b} but they are not well represented in low-redshift quasar samples, making it challenging to reconstruct them using standard techniques (see discussions in \citealt{davies2018a} and \citealt{greig2019}).
 Here we model the quasar's emission by creating composite spectra based on spectra of the SDSS DR12 quasar catalog \citep{paris2017} with comparable \civ\ properties to P183+05  \citep[see][]{mortlock2011,bosman2015,banados2018a}. Our approach can be summarized in the following steps.

 \begin{enumerate}
\item We select quasars from the SDSS DR12 catalog flagged as non broad-absorption line quasars in the redshift range $2.1<z<2.4$ (i.e., quasar spectra covering the $\lya$, \civ, and \mgii\ lines).  To preselect quasars with extreme blueshifts we require the pipeline redshift to be blueshifted by at least $1500\, \kms$ from the \mgii-derived redshift. This yields 3851 quasars.

\item We then measure the median S/N at the continuum level of the \civ\ region for each spectrum. We only retain objects with a median S/N$>5$ in the wavelength range $1450-1500\,$\AA. After this step, we are left with 2145 quasars.

\item We model the \civ\ line wavelength region for each quasar as a power law plus a Gaussian. We estimate the \civ\ EWs and their velocity offsets with respect to the \mgii\ redshift, which is thought to be a reliable systemic redshift estimator \citep{richards2002a}.

\item
We require the quasars to have \civ\ EWs consistent with that measured in P183+05 (EW$=11.8\pm 0.7\,$\AA) at the 3$\sigma$ level.  Given that some $z>6$ quasars have significant blueshifts between \mgii\ and \cii\ lines \citep{venemans2016}, we select SDSS quasars with a \civ\ blueshift (measured from the \mgii\ line) consistent within $1000\,\kms$ of the P183+05 \civ\ blueshift ($5057 \pm 93\, \kms$; measured from the \cii\ line). After applying these criteria, we are left with 61 P183+05 ``analogs.''

\item We fit the continua of these analogs by a slow-varying spline to remove strong absorption systems and noisy regions. We then normalize each spectrum at $1290\,$\AA\ and average them. This mean composite spectrum and the $\pm 1\sigma$ dispersion around the mean are shown as the red line and orange region in Figure \ref{fig:sdssmatch}, respectively. The dispersion at $1290\,$\AA\ is zero by construction and increases at shorter and longer wavelengths. The mean spectrum (red line) matches the general features in the observed spectrum of P183+05 well and predicts a weaker $\lya$ line than observed in typical low-redshift quasars \citep{paris2011}.
\end{enumerate}

Using this continuum model, we fit a Voigt profile to the damping wing with a fixed centroid at $z_{\rm DLA}=6.40392$,  the redshift of the DLA derived from low-excitation metal lines (Section \ref{sec:dla-metals}).
A least-squares optimization yields a best-fit value of $N_{\mathrm{HI}}=10^{20.77}\,$cm$^{-2}$, which reproduces the data  well and is consistent with no emission spikes at the expected location of the Ly$\beta$ absorption (see Figure \ref{fig:fors2}). We assume a conservative uncertainty of 0.25 dex, which represents the credible range allowed by the S/N of our spectrum.
This was determined by overplotting Voigt profiles, varying the input column density to identify the range allowed by the data\footnote{
A subjective visual measurement of the uncertainty of the absorption profile like the one employed here is the standard methodology in the field \cite[e.g.,][]{rafelski2012,selsing2019} because simple $\chi^2$ minimization yields unrealistically small errors.
}. We note that the least-squares best-fit Voigt profile ($N_{\mathrm{HI}}=10^{20.77}\,$cm$^{-2}$) seems systematically lower than the data at $\lambda_{\rm rest}>1216\,$\AA. Even though it is possible to find profiles that better match the data at those wavelengths by lowering $N_{\mathrm{HI}}$, in those cases the match to the data at  $\lambda_{\rm rest}<1216\,$\AA\ worsens (see, e.g., the violet dashed lines in the top panel of Figure \ref{fig:dlaigm}).

 A closer look at the top panel of Figure \ref{fig:dlaigm} reveals that a DLA alone cannot reproduce the sharp drop in flux around $\lambda_{\rm rest}=1212-1215\,$\AA\ (see yellow region in the residuals panel). We attempted other DLA fits using alternative continua as intrinsic model in Appendix \ref{app:alternative} but were not able to find another case that could fit the data better. On the other hand, a damping wing produced by a $\fhi=0.80$ IGM can reproduce the data around $\lambda_{\rm rest}=1212-1215\,$\AA\ well (see Appendix \ref{app:igmpos}). This indicates that the combined effects of a neutral IGM and a DLA might be the most plausible scenario.
 In Appendix \ref{app:igmpdla} we discuss that the joint IGM+DLA fit is degenerate (see Figure \ref{fig:app_igm} and \ref{fig:app_dlapars}). We take as our preferred model the combined fit shown in the bottom panel of  Figure \ref{fig:dlaigm}, i.e., we fix a $\fhi=0.10$ IGM contribution and then we find a best-fit DLA with $N_{\mathrm{HI}}=10^{20.68\pm 0.25}\,$cm$^{-2}$. This is motivated by the following two facts: (i) we see metal absorption lines (Figure \ref{fig:metals}) with low internal velocity dispersion of the gas (narrow lines) that are characteristic of very metal-poor DLAs \citep[e.g.,][]{cooke2015} and (ii) IGM neutral fractions $\fhi\gtrsim 0.4$ have only been reported at significantly higher redshifts \citep[$z\sim 7.5$; e.g.,][]{davies2018b,hoag2019}. We note, however, that joint fits with an IGM neutral fraction in the range $\fhi=0.05-0.50$ produce  fits to the data comparable to our assumed scenario. A different way to reproduce the data in the $\lambda_{\rm rest}=1212-1215\,$\AA\ region is to include additional saturated but weak absorbers in the proximity zone of the quasar. For example, an additional absorber with  $\log N_{\mathrm{HI}}= 15.51$ at $z=6.427$ is consistent with the sharp drop in flux at $\sim 1214\,$\AA. {When including the $z=6.427$ absorber the corresponding best-fit $\log N_{\mathrm{HI}}$ for the DLA is $20.68$}.
 Thus a $\fhi=0$ scenario cannot be completely ruled out with the current data.
 Our assumed fiducial value for the DLA column density with its conservative uncertainty of 0.25 dex, includes all the $N_{\mathrm{HI}}$ best-fitting values when considering the contribution of an IGM with a neutral fraction in the range $\fhi=0-0.38$ (see Figure \ref{fig:app_dlapars}).

\subsection{Ionization and dust corrections}
\label{sec:ionization}

For systems with large column densities of hydrogen like DLAs in which nearly all of the metals are in either their neutral or first excited states (as is the case for the system of this study) the ionization corrections are negligible \citep{vladilo2001}. However, because of the proximity of the DLA to P183+05 it is important to quantify whether the ionization radiation from the quasar would significantly affect our derived metallicities.
We use the spectral synthesis code Cloudy \citep{ferland2017} to explore the maximum ionization corrections allowed by the observed upper limit on $\log N_{\mathrm{SiIV}}-\log N_{\mathrm{SiII}}<-0.53$.  The simulation was made to simultaneously reproduce the observed neutral hydrogen column density $\log N_{\mathrm{HI}}=20.68$ and the observed upper limit on the \siiv/\siii\ ratio. This can be achieved by placing an active galactic nucleus source of the same observed
1450\AA\ brightness as P183+05 at $>375$ physical kpc from the DLA cloud. A detection of \siiv\ at this level would indeed require two thirds of the hydrogen of the DLA to be ionized ($N_{\rm HI}/N_{\rm H}>0.33$), leading to a total hydrogen column density of $\log N_{\mathrm{H}}=21.16$.
Nevertheless, we find that the correction is negligible for $N_{\rm OI}/N_{\rm HI}$, which was expected because of the charge-exchange equilibrium between \oi\ and \hi. The other common low ion abundances (\siii/\hi, \ciib/\hi, and \mgii/\hi) are also remarkably insensitive to the neutral gas fraction (\hi/H); the required correction factors for the metal abundances would be as small as  $(N_{\mathrm{SiII}}/N_{\mathrm{HI}})/(N(\mathrm{Si})/N(\mathrm{H}))=1.16$,
$(N_{\mathrm{CII}}/N_{\mathrm{HI}})/(N_{\mathrm{C}}/N_{\mathrm{H}})=1.09$, and $(N_{\mathrm{MgII}}/N_{\mathrm{HI}})/(N_{\mathrm{Mg}}/$ $N_{\mathrm{H}})=0.95$. As \siiv\ is not actually detected, the actual corrections would be even smaller. Based on these results, and the fact that the observed metallicities are very similar to that based on \oi/\hi, we do not apply ionization corrections. In this way are also able to directly compare our results to similar high-redshift absorption systems from the literature that do not include ionization corrections \citep[e.g.,][]{beckerG2012,dodorico2018}.

Another factor to consider is dust depletion, which might have fundamental consequences for derived metallicities as we know that large amounts of dust exist even in some of the most distant galaxies identified to date \citep[e.g., ][]{venemans2018,tamura2019}. Indeed, recent studies have demonstrated that dust-depletion corrections are important for the derived metallicities of high-redshift DLAs associated with both GRBs and quasars \citep[e.g.,][]{bolmer2019,decia2018,poudel2018} although it is also well established that the levels of depletion
reduce with decreasing metallicity \citep{kulkarni2015}.
 Thus, for our metallicity estimate we use the element oxygen which is very weakly affected by both dust depletion and ionization corrections.

\floattable
\begin{deluxetable}{lCCCCC}
\tablenum{1}
\label{tab:ews}
\tablecaption{Properties of the absorption lines identified in the DLA at $z=6.40392$ toward the quasar P183+05 at $z=6.4386$.}
\tablewidth{0pt}
\tablehead{
\colhead{Line ID} &  \colhead{$\lambda_{\rm rest}$} &  \colhead{EW$_{\rm rest}$} & \colhead{$\log\, N_{\rm X}$\tablenotemark{a}} & \colhead{$v_{\rm min},v_{\rm max}$\tablenotemark{b}}  \\
\colhead{} & \colhead{(\AA)} & \colhead{(\AA)} & \colhead{(cm$^{-2}$)} & \colhead{($\rm km~s^{-1}$)} & \colhead{}
}
\startdata
\ciib  & 1334.5323  & 0.30\pm 0.03  & 14.30\pm0.05 &  -105,105     \\
\civ  & 1548.204  & <0.14  & <13.54 & -150,150     \\
\civ  & 1550.781  & <0.09  & <13.67 & -150,150     \\
\nv  & 1238.821  & <0.06  & <13.47 & -150,150     \\
\nv  & 1242.804  & <0.05  & <13.79 & -150,150     \\
\oi    & 1302.1685  & 0.16 \pm 0.02 & 14.45\pm 0.20\, (0.06) & -75,80  \\
\mgii   & 2796.3543  & 0.70 \pm 0.05 & 13.37\pm0.04 & -150,150           \\
\mgii   & 2803.5315  & 0.39 \pm 0.05 & 13.38\pm0.04 & -150,150   \\
\alii  & 1670.7886  & 0.12 \pm 0.01 & 12.49\pm 0.20\, (0.06) & -70,70   \\
\siii\tablenotemark{c}   & 1260.4221  & 0.38 \pm 0.03 & 13.53\pm0.04 & -150,150            \\
\siii\tablenotemark{c}    & 1304.3702  & 0.04 \pm 0.01 & 13.54\pm0.16 & -150,150            \\
\siii\tablenotemark{c}    & 1526.7070  & 0.29 \pm 0.03 & 14.15\pm0.05 & -150,150            \\
\siiv    & 1393.7602  & <0.08 & <13.0 & -150,150            \\
\siiv    & 1402.7729  & <0.07 & <13.19 & -150,150            \\
\feii  & 2382.7652  & 0.22 \pm 0.03 & 13.19\pm0.05 & -115,90  \\
\enddata
\tablenotetext{a}{For \oi\ and \alii\ the uncertainty in parenthesis is from the AODM but the assumed uncertainties are more conservative to include potentially hidden saturation at the resolution of our data (see Appendix \ref{app:saturation}).}
\tablenotetext{b}{Minimum and maximum velocities with respect to the DLA's velocity centroid used by the AODM to measure the column densities.}
\tablenotetext{c} {We consider the column density of \siii\ $\lambda 1260$ as an upper limit as it is potentially contaminated with a \civ\ $\lambda 1548$ absorption line from a system at $z=5.0172$. The column density of \siii\ $\lambda 1526$ overpredicts the other transitions. See the discussion in Appendix \ref{app:saturation} for details.}
\tablecomments{All reported limits correspond to $3\sigma$. We note that the EW significance obtained for \civ\ $\lambda 1548$, \siiv $\lambda1393$, and  \nv\ $\lambda 1242$ are $3.7\sigma$, $5.3\sigma$, and $6.4\sigma$ respectively. However, we report them as $3\sigma$ limits as their putative detections are not convincing (see Figure \ref{fig:metals}).}
\end{deluxetable}

\floattable
\begin{deluxetable}{lCCCCCC}
\tablenum{2}
\label{tab:abundances}
\tablecaption{Column densities and relative abundances of the DLA at $z=6.40392$ toward the quasar P183+05 at $z=6.4386$}
\tablewidth{0pt}
\tablehead{
\colhead{X} & \colhead{$\log \epsilon (X)_\odot$\tablenotemark{a}} & \colhead{$\log\, N_{\rm X}$} & \colhead{[X/H]} & \colhead{[X/O]} & \colhead{[X/Si]} & \colhead{[X/Fe]} 
}
\startdata
H  & 12.00 & 20.68\pm0.25   &  -             & 2.92\pm0.32 & >2.66   & 2.94\pm0.26   \\
C  & 8.43 & 14.30\pm0.05    &  -2.81\pm0.26   & 0.11\pm0.21 & >-0.15 & 0.13\pm0.07\\
O  & 8.69 & 14.45\pm0.20    &  -2.92\pm0.32   & -           & >-0.26 & 0.02\pm0.21 \\
Mg & 7.53 & 13.37 \pm 0.03  &  -2.84\pm 0.25  & 0.09\pm0.20 & >-0.18 & 0.10\pm0.06 \\
Al & 6.43 & 12.49\pm0.20    & -2.62\pm0.32    & 0.31\pm0.28 & >0.04  & 0.32\pm0.21 \\
Si\tablenotemark{b}  & 7.51 & <13.53 & <-2.66 & <0.26       & -      & <0.28 \\
Fe & 7.45 & 13.19\pm 0.05   & -2.94\pm0.26    & -0.02\pm0.21 & >-0.28 & - \\
\enddata
\tablenotetext{a} {$\log \epsilon (X)_\odot = 12 + \log(N_{\rm X}/N_{\rm H})_\odot$.  The solar abundances are taken from the meteoritic values reported in \cite{asplund2009}, except for C and O, for which we use the photospheric values.}
\tablenotetext{b} {Here we only used limits from \siii\ $\lambda 1260$, which are more stringent.  }
\tablecomments{The column densities $N_{\rm X}$ are in units of cm$^{-2}$.  Relative abundances do not include ionization or depletion corrections.
Our assumed metallicity is based on oxygen ([O/H]$=-2.92\pm0.32$) as is an undepleted element and has an ionization potential similar to \hi. Element ratios are relative to the solar values, i.e., $[{\rm X/Y}] = \log(N_{\rm X}/N_{\rm Y}) - \log(N_{\rm X}/N_{\rm Y})_\odot$. }
\end{deluxetable}

\section{Discussion} \label{sec:discussion}

\subsection{Metal-poor DLA}

The derived metallicity of [O/H]$=-2.92\pm 0.32$ makes this object one of the most metal-poor DLAs currently known. In fact, the three most metal poor DLAs were reported recently: \cite{cooke2016,cooke2017} presented two $z\sim 3$ DLAs with metallicities of [O/H]$=-2.804\pm 0.015$  and  [O/H]$=-3.05\pm 0.05$ while  \cite{dodorico2018} reported a $z=5.94$ DLA with  metallicity [O/H]$\geq -2.9$ (and [Si/H]$=-2.86\pm 0.14$, but not dust-corrected). Thus, the metallicity of the DLA of this paper is consistent with the current record holders, within the uncertainties.
We note that the fact that this $z=6.4$ DLA and the $z=5.94$ DLA from \cite{dodorico2018} do not present evidence of high-excitation ions (e.g., \civ) departs from typical DLAs at lower redshift ($z\sim 2-3$), which tend to have associated \civ\ (e.g., \citealt{rubin2015}), but is not unexpected given the diminishing rate of incidence of high-ionization absorbers beyond $z\sim 5$ (e.g., \citealt{beckerG2009,codoreanu2018,cooper2019}).
This is consistent with the idea that the gas causing \civ\ and higher ions resides in and follows the density evolution of the general IGM, while the DLA gas is more closely associated with galactic halos \citep{rauch1997}.

\begin{figure*}[!ht]
\centering
\figurenum{5}
\plotone{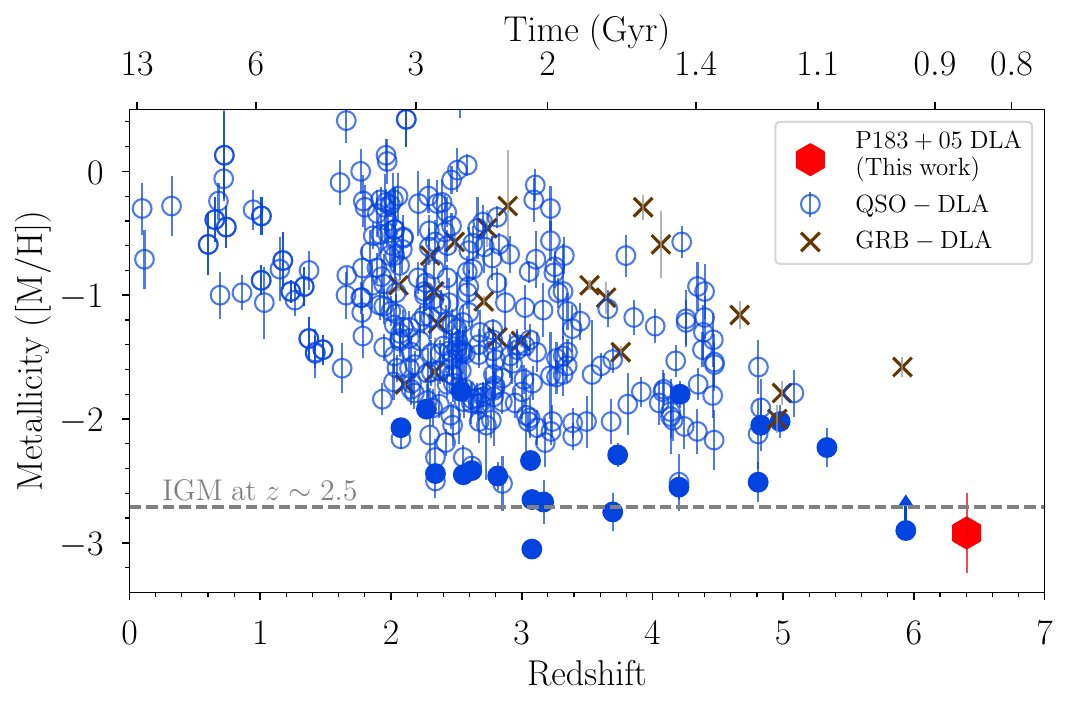}
\caption{
 DLA metallicity ([M/H]) vs. redshift. The top axis shows the cosmic time from the big bang. The blue open circles are the dust-corrected metallicities compiled by \cite{decia2016,decia2018}.
 The blue filled circles are metallicities using the undepleted element oxygen from the compilation of the most metal-poor DLAs known by \cite{cooke2017}, the $z>4.5$ absorbers reported by \cite{poudel2018}, and the recent $z=5.94$ PDLA identified by \cite{dodorico2018}.
 For completeness, we  show the dust-corrected metallicities of the GRB-DLAs reported by \cite{bolmer2019} as brown crosses. For comparison, the horizontal dashed line represents the mean IGM metallicity found at $z\sim 2.5$  ([O/H]$=-2.71$; \citealt{simcoe2004}; we increased the reported metallicity by 0.14 dex to match the oxygen solar abundance assumed here, see Table \ref{tab:abundances}).  The PDLA discovered in this paper (red hexagon) is among the most metal-poor systems known, with a metallicity comparable to that of the IGM at lower redshifts.
}
\label{fig:metal_vs_z}
 \end{figure*}

In Figure \ref{fig:metal_vs_z} we show metallicity versus redshift for the DLAs compiled by \cite{decia2018}, the metal-poor DLAs compiled by \cite{cooke2017}, the $z>4.5$ absorbers reported by \cite{poudel2018}, and the $z=5.94$ DLA reported by \cite{dodorico2018}.  All the metallicities are either dust-corrected or based on the undepleted element oxygen.
Our new data point at $z=6.40392$ (red hexagon) and the $z=5.94$ DLA are in line with the general trend of decreasing metallicity with redshift.

It is remarkable that the metallicities of the two highest-redshift DLAs still do not drop significantly below other measurements of the metallicity of overdense gas, but simply straddle the lower-metallicity envelope prescribed by the mean metallicity of the IGM (Figure 4), which is nearly constant over a long-redshift baseline ranging from $z\sim 2.5$ \citep[e.g., \civ\ and \ovi\ absorbers:][]{rauch1997,schaye2003,simcoe2004}  through $5<z<6$ \citep[e.g., \oi\ systems;][]{keating2014}, to the present DLA at $z>6$. Modeling of what appears to be a metallicity floor in the IGM by contemporary outflows without invoking even more ancient phases of pre-enrichment has remained a challenge both at intermediate \citep{kawata2007} and high redshift \citep{keating2016}.

\subsection{Chemical enrichment}

This DLA is seen only $\sim$$850\,$Myr after the big bang, thus there was not much time for metal enrichment. This is in line with it being one of the most metal-poor DLAs currently known, as discussed above. Therefore, this makes it an interesting system to ask whether its chemical abundance patterns could be explained by the yields of the first metal-free stars.

\begin{figure*}
\figurenum{6}
\plotone{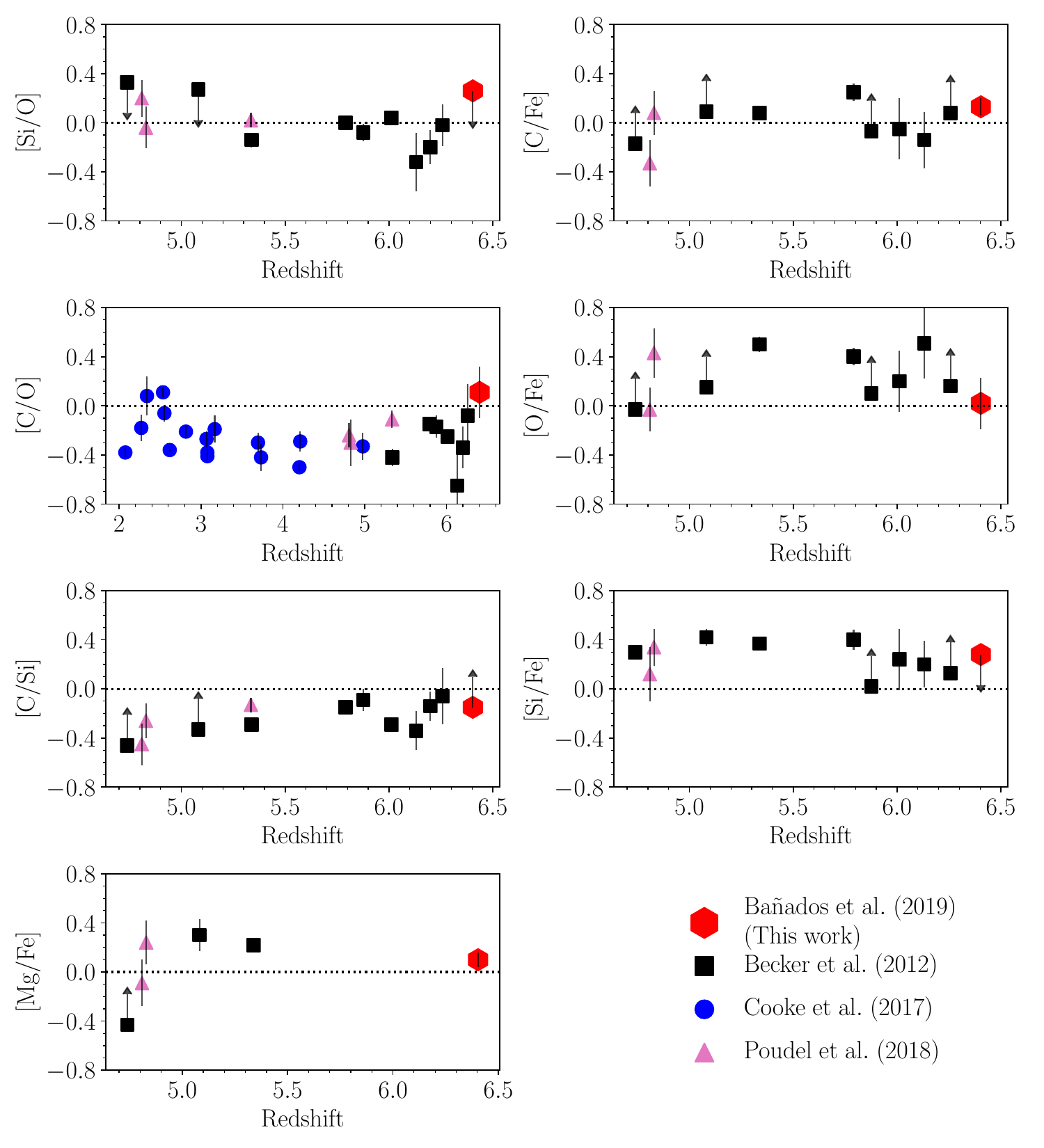}
\caption{Relative abundances of the DLA identified in this paper at $z=6.40392$ (red hexagon) compared to data from the literature. The data points from \cite{beckerG2012} 
are shown as black squares. The blue circles show the compilation of the most metal-poor DLAs known by \cite{cooke2017}, and the $z>4.5$ absorbers reported by \cite{poudel2018} are shown as pink triangles. \label{fig:beckercomp}}
\end{figure*}

We note that, in general, the element ratios of this system are close to solar (see Table \ref{tab:abundances}), which differs from the sub-solar abundances observed in typical metal-poor DLAs at $z\sim 3$ (see Figure 13 in \citealt{cooke2011b}). Nevertheless, the solar relative abundances of this system are consistent with the patterns observed by \cite{beckerG2012} and \cite{poudel2018} in a sample of absorbers at higher redshifts (Figure \ref{fig:beckercomp}).

\begin{figure*}
\figurenum{7}
\plotone{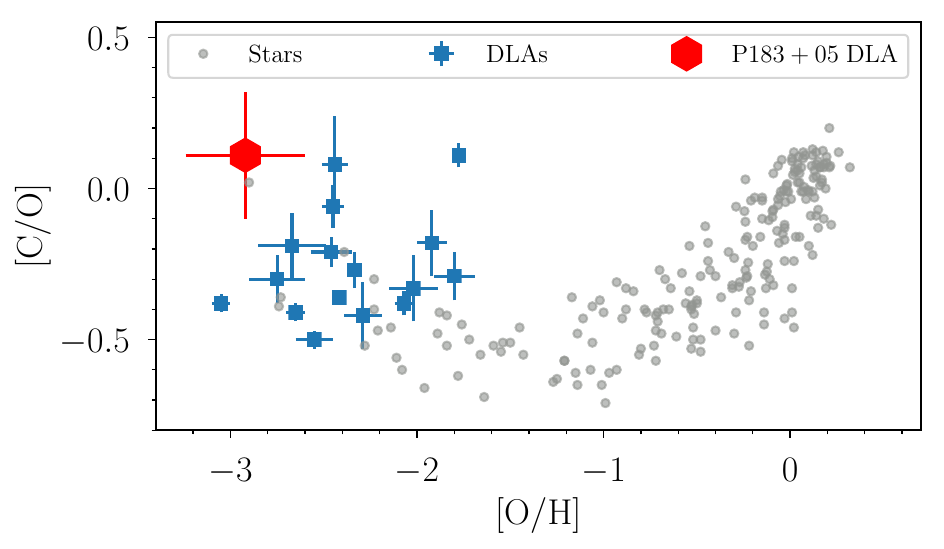}
\caption{Chemical evolution of [C/O] vs. metallicity [O/H]; adapted from Figure 5 of \cite{cooke2017}. The gray circles are measurements for stars (\citealt{bensby2006,fabbian2009,nissen2014}), the blue squares for metal-poor DLAs \citep{cooke2017}, while the red hexagon is the new measurement for the $z=6.40392$ DLA toward P183+05 presented in this work. The increase of [C/O] for metallicities [O/H]$<-1$ for both stars and DLAs is not yet fully understood (see text). The DLA presented here follows this trend.  \label{fig:cooke}}
\end{figure*}

The [C/O] abundance has received significant attention over recent years as it still challenges our understanding of stellar nucleosynthesis. [C/O] increases linearly with metallicity from $\sim$$-0.5$ to $\sim$$0.5$ when [O/H]$>-1$. This has been explained by the yields of carbon produced by massive rotating stars, which increase with metallicity, in addition to a delayed contribution of carbon from lower-mass stars \citep{akerman2004}. Conversely, current models of Population II nucleosynthesis predict that [C/O] should decrease or reach a plateau below a metallicity of [O/H]$\sim$$-1$. However, both observations of metal-poor stars \citep{akerman2004,fabbian2009} and metal-poor DLAs \citep{cooke2017} show the opposite trend, i.e., [C/O] increases to solar abundance at lower metallicities (see Figure \ref{fig:cooke}).    This intriguing trend has been interpreted as an enhanced production of carbon by Population III stars or by rapidly rotating Population II stars \citep{cooke2017}.
 On the other hand, recent simulations show that high values of [C/O] can be due to the enrichment by asymptotic giant branch stars formed before $z=6$ without including Population III stars \citep{sharma2018}.
Our new data point follows and further expands
this empirical tendency (see red hexagon in Figure \ref{fig:cooke}), which is still lacking a definite explanation.

To see what possible formation scenarios could describe the chemical composition of the DLA of this paper we refer to the discussions in \cite{cooke2011b} and \cite{ma2017b}, with special attention to their Figures 14 and 5, respectively.  Both studies investigate the expected abundance patterns produced by the first stars in high-redshift DLAs.
The relative abundances observed in the DLA toward P183+05 do not match any of the Population III patterns expected by these studies. In particular, none reproduces the [C/O] abundance.

To investigate whether there are nucleosynthesis models of massive metal-free stars that can reproduce the observed abundances seen in the DLA of P183+05 we fit Population III supernova yields \citep{heger2010} using the tools developed\footnote{\url{https://github.com/alexji/alexmods/alex_starfit.py}} by \citealt[][]{frebel2019} (see their section 4.4). The best fits have a very similar $\chi^2$ and seem almost indistinguishable. For better visualization of the different models, in Figure \ref{fig:starfit} we show the first, 10th, 20th, and 30th best-fitting models (i.e., the model with the lowest $\chi^2$, the 10th lowest $\chi^2$, etc.).
The best-fitting models prefer a progenitor mass in the range of $13\, M_\odot\lesssim M \lesssim 15\, M_\odot$ and a range of explosion energy.  We note that these models give the yields produced by a single-progenitor supernovae and it is  clear that they cannot reproduce all the abundances observed in the DLA, particularly the Al abundance. This could indicate that more than one supernova was responsible for the enrichment in the DLA, which could hide any potential signature due to Population III stars \citep[see also ][]{maio2015}.   In addition, we do not observe the level of carbon enhancement ([C/Fe]$>0.70$) seen in most of the very metal-poor stars (see e.g., \citealt{placco2014}), which is thought to be a signature produced by the first stars.

Therefore, we do not find evidence that the yields of Population III stars need to be invoked to explain the chemical enrichment of the DLA presented here.

\section{Summary} \label{sec:summary}

We identify a strong Ly$\alpha$ damping wing profile in the spectrum of the $z=6.4386$ quasar P183+05 in addition to several narrow metal absorption lines at $z=6.40392$ (Figure \ref{fig:metals}). We find that the best explanation for the absorption profile near the Ly$\alpha$ region is either a combination of a DLA and a $\fhi=0.05-0.38$ IGM in the surroundings of the quasar or a DLA plus additional weaker absorbers in the quasar's proximity zone (Figure \ref{fig:dlaigm}). This DLA is remarkable for several reasons.
\begin{itemize}

\item It is currently the most distant absorption system known ($z=6.40392$; i.e., only $857\,$Myr after the big bang) where a direct metallicity estimation is possible.

\item It is among the most metal-poor DLAs currently known with a metallicity of [O/H]$=-2.92\pm 0.32$ (i.e., $\sim$$1/800$ times the solar value). This metallicity is consistent with that of the current most metal-poor DLAs known and with the mean metallicity of the IGM measured at much lower redshifts (see Figure \ref{fig:metal_vs_z}).

\item It has chemical abundance patterns that do not match the expected yields of Population III stars. Thus, we do not find evidence of metal enrichment produced by the first stars in this high-redshift DLA, seen when the universe was 6\% of its present age.

\end{itemize}

 Absorption systems toward the highest-redshift quasars like that presented here would be ideal targets for  high-resolution ($R\sim 50,000 - 100,000$) near-infrared spectroscopy with the instruments being planned for the next generation of 25--40 m telescopes \citep[e.g.,][]{zerbi2014,jaffe2016}. Therefore, finding more of these systems could greatly enlighten our understanding of the epoch of reionization and the formation of the first stars.

\begin{figure*}
\figurenum{8}
\plotone{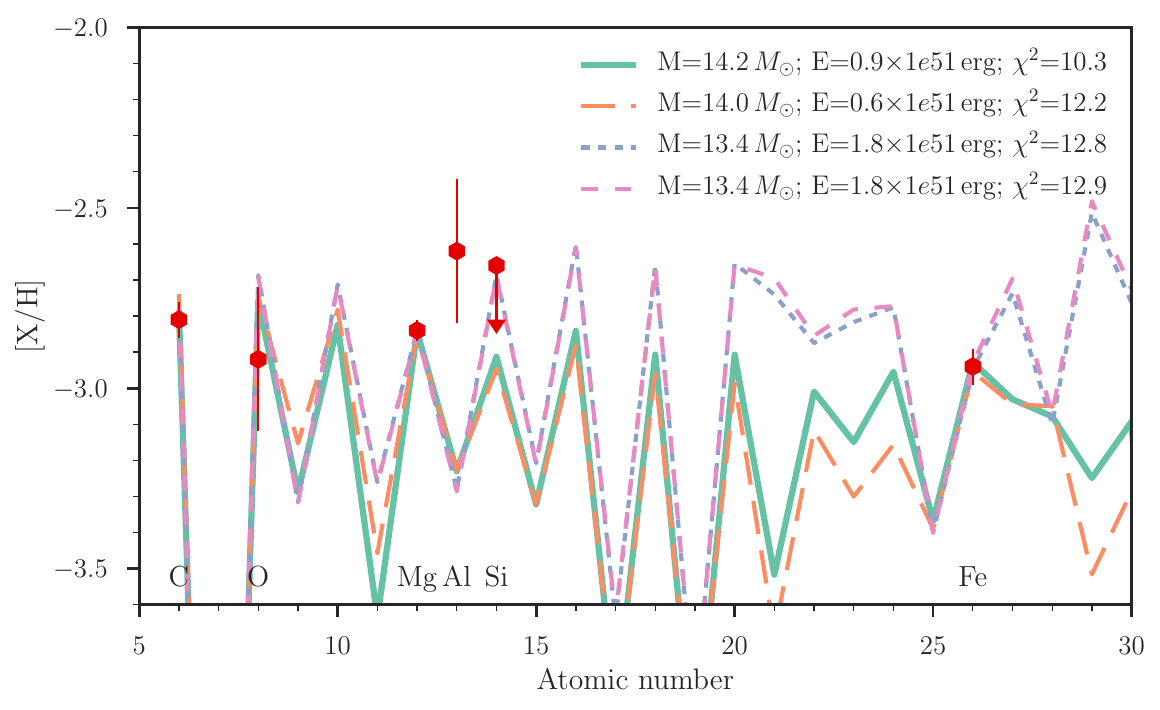}
\caption{
First (lowest $\chi^2$), 10th, 20th, and 30th best-fitting metal-free supernova yield models from \cite{heger2010} (lines, ranked by $\chi^2$ in the legend) that best match the abundance pattern of the DLA proximate to P183+05 (red hexagons).
  We do not find a model that can simultaneously reproduce all the abundances observed in the present DLA (e.g., see aluminum in this figure). Note that the uncertainties used and shown here do not include the error on $\log N(\mathrm{H})$, which are not relevant for comparing the relative element abundances to the model predictions.
 }
\label{fig:starfit}
\end{figure*}

\vspace{5mm}
\facilities{Magellan:Baade (FIRE), VLT:Antu (FORS2)}
\software{Astropy \citep{astropy2018},
Matplotlib \citep[][\url{http://www.matplotlib.org}]{hunter2007},
Cloudy \citep{ferland2017},
pyigm (\url{https://github.com/pyigm/pyigm}),
linetools (\url{https://github.com/linetools})}

\acknowledgments
We thank the referee for insightful and constructive suggestions that have substantially improved this manuscript. We thank Max Pettini for reading a previous version of this manuscript, providing valuable comments.
We thank Ryan Cooke, Alex Ji, Andrew McWilliam, Ian Roderer, Gwen Rudie, and Kevin Schlaufman for insightful conversations that helped to shape this work.
 We thank Ryan Cooke for providing the data to produce Figure \ref{fig:cooke}. We thank Alex Ji for sharing the codes necessary to produce Figure \ref{fig:starfit}.
E.P.F., B.P.V, and F.W. acknowledge funding through the ERC grant ``Cosmic Dawn.''
S.C. is supported by the National Science Foundation Graduate Research Fellowship under Grant No. DGE 1106400 and gratefully acknowledges direct funding from the MIT Undergraduate Research Opportunity program (UROP).
 This work is based on data collected with the Magellan Baade telescope located at Las Campanas Observatory, Chile.
 Based on observations collected at the European Southern Observatory under ESO program 095.A-0375(A).
 This research has made use of NASA's Astrophysics Data System.

%



\appendix
\section{Exploring the IGM/DLA degeneracy}
\label{app:igmpos}

As discussed in the main text, without any further information the absorption profile seen in the spectrum of P183+05 could be caused by a neutral IGM or a DLA. While we know that there must be an absorber in front of P183+05 given the narrow metal absorption lines seen in its spectrum (Figure \ref{fig:metals}), an absorption profile caused only by a DLA does not provide a satisfactory fit to the data (Figure \ref{fig:fors2}).
Here we explore different continuum models and the effects of combined fits of a neutral IGM and a DLA.

\subsection{Alternative continuum models}
\label{app:alternative}
\begin{figure}
\centering
\figurenum{9}
\plotone{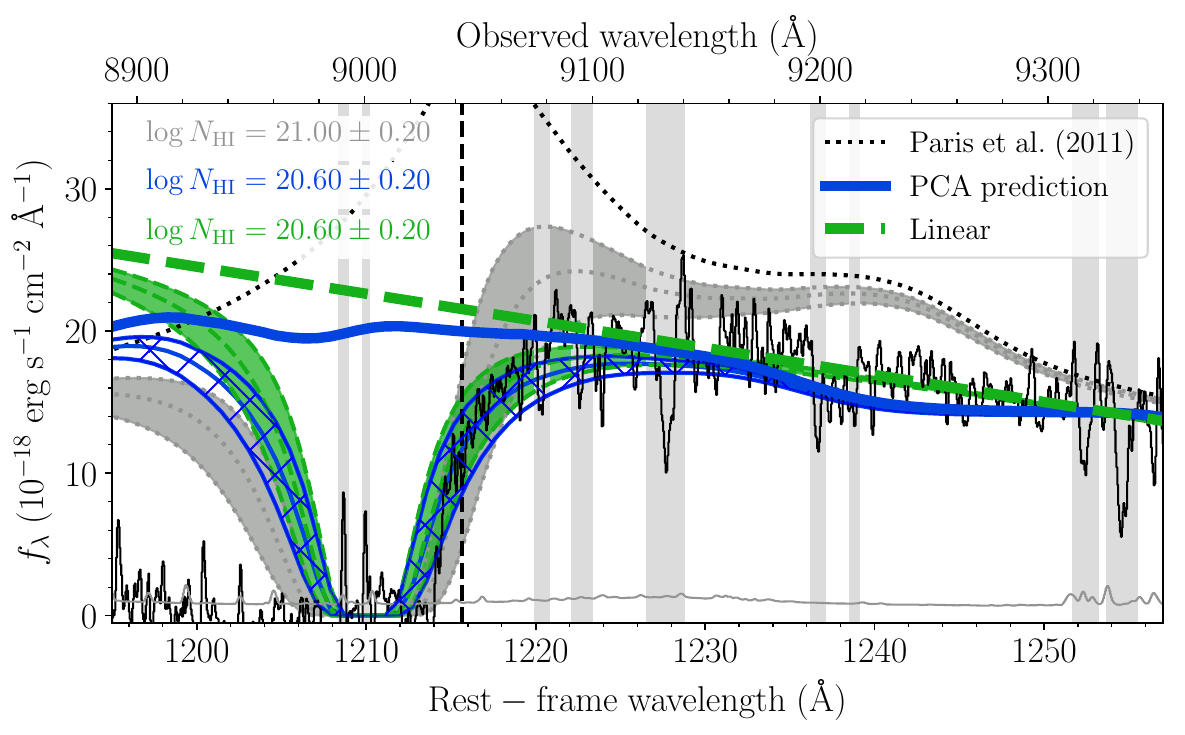}
\caption{
Same spectrum as in the top panel of Figure \ref{fig:dlaigm} but this time showing DLA profiles for three alternative continuum models. The dotted line is the mean SDSS quasar from \cite{paris2011}, the thick blue line is the PCA prediction obtained using the methodology of \cite{davies2018a}, while the green dashed line is an unrealistic/simplistic linear model. None of the alternative models produces a satisfactory match to the data.
}
\label{fig:app_dlaalt}
 \end{figure}

We explore the results of using three alternative continuum models of the $\lya$ region as shown in Figure \ref{fig:app_dlaalt}.
First, we use the mean SDSS quasar spectrum from \cite{paris2011} as a model. The mean SDSS quasar has stronger emission lines than P183+05 (see also Figure \ref{fig:sdssmatch}) and therefore  requires a higher neutral hydrogen column density than our fiducial model to match the onset of the damping wing. However, given the expected stronger \nv\ line is not possible to find a reasonable DLA fit using this continuum model to match the data (see dotted line and shaded gray region in Figure \ref{fig:app_dlaalt}).  Second, we reconstruct the
 $\lya$ region using the method presented in \cite{davies2018a}. This method is based on PCA decomposition and is specially designed to reconstruct the $\lya$ region of $z\gtrsim 6$ quasars. The PCA continuum model is shown as a solid blue line in Figures \ref{fig:sdssmatch} and \ref{fig:app_dlaalt}. Although the PCA continuum matches well the general characteristics of P183+05, it clearly underpredicts the observed flux near the $\lya$ region and it is virtually impossible to fit a Voigt profile consistent with the data using this continuum model (see solid blue line and hatched region in Figure \ref{fig:app_dlaalt}). Possible explanations are that the modest S/N of our data hinders this method and that we do not cover crucial emission lines (e.g., \ciii) that have strong predictive power  (similar effects would impact other sophisticated methods that take advantage of high-S/N spectra and availability of several key emission lines, e.g., \citealt{greig2017b}).
 Third, we use a rather simplistic and unrealistic $\lya$ model consisting of a linear fit to the data right redward of the $\lya$ line (green dashed line in Figure \ref{fig:app_dlaalt}).  This is to showcase an extreme scenario with a very weak $\lya$ line.  The last, unrealistic case gives a more reasonable DLA fit than the other two alternative continua but it is still far from being a good match to the data.

\subsection{IGM + DLA joint fits}
\label{app:igmpdla}

Here we explore different alternatives, combining the effects of a neutral IGM and a DLA.
 We model the IGM damping wing following the formalism of \cite{miralda-escude1998}, assuming a constant IGM neutral fraction between the quasar's proximity zone and $z = 6$, while being completely ionized at $z < 6$.  The proximity zone is typically defined as the physical radius at which the transmission drops to 10\%, which for P183+05 corresponds to 0.67 physical Mpc. We note that this proximity zone is much smaller than the expected $\sim 4.5$ physical Mpc for a quasar with the luminosity and redshift of P183+05 \citep{eilers2017}. In this particular case a small proximity zone does not come as a surprise given the existence of the proximate DLA, which is in fact one of the possible explanations for the small proximity zones found by \cite{eilers2017}.

\begin{figure}
\centering
\figurenum{10}
\plotone{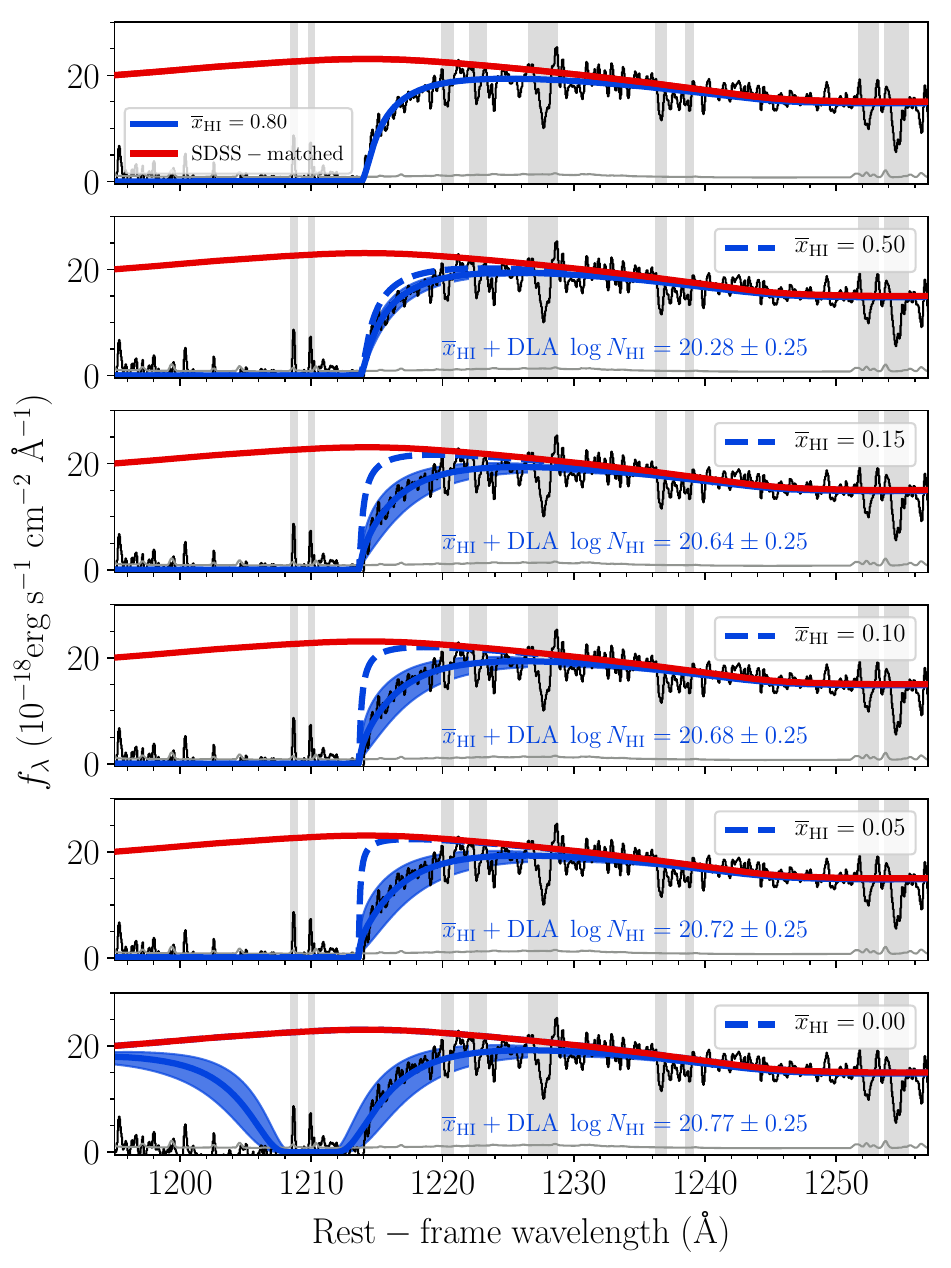}
\caption{
All panels: FIRE spectrum of P183+05 (black line), its 1$\sigma$ error vector (gray line), and the SDSS-matched composite spectrum (red line). In the top panel we show the best-fit IGM damping wing profile, with  $\fhi=0.80$ in the surrounding of the quasar (blue line). The rest of the panels show the attenuation caused by an IGM (dashed blue lines, see legends) with $\fhi=0.5$, $\fhi=0.15$, $\fhi=0.10$,  $\fhi=0.05$, and $\fhi=0.00$ (i.e., no IGM contribution in the  bottom panel). The solid blue lines represent the best-fit DLA model using as input the continuum already damped by the IGM (dashed lines). The blue regions show the effect of varying the best-fit $N_{\mathrm{HI}}$ by 0.25 dex.
}
\label{fig:app_igm}
 \end{figure}

\begin{figure}
\centering
\figurenum{11}
\plotone{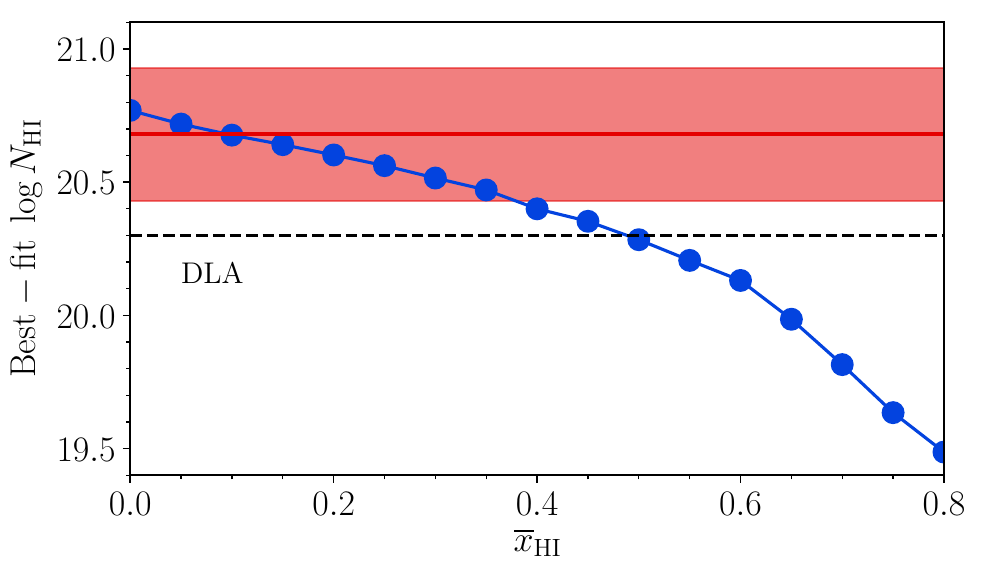}
\caption{
Best-fit $\log N_{\mathrm{HI}}$  for a DLA model applied to the continuum model attenuated by an IGM with a neutral fraction $\fhi$ (see also Figure \ref{fig:app_igm}).
Our fiducial value is $\log N_{\mathrm{HI}}=20.68 \pm 0.25$ (red line and shaded region), which corresponds to the best-fitting value when $\fhi=0.10$. The conservative uncertainty encompasses the column densities allowed for all cases with an IGM $\fhi<0.38$. The dashed line marks the column density that defines DLAs: $N_{\mathrm{HI}} > 2\times 10^{20}\,$cm$^{-2}$.
}
\label{fig:app_dlapars}
 \end{figure}

For simplicity, in all cases here we use the SDSS-matched spectrum as intrinsic continuum of the quasar. If we fit the data with a combined model of IGM+DLA with three free parameters ($\fhi$, $N_{\mathrm{HI}}$, and the proximity zone), the outcome is highly degenerate. We use the differential evolution algorithm \citep{storn1997} to find the global minimum of the root-mean-deviation between the data and the model. This yields $\fhi=0.81$, $\log N_{\mathrm{HI}}=19.55$, and a proximity zone of 0.72 physical Mpc. The global minimum is dominated by a very neutral IGM  because an IGM absorption can reproduce the steep step in flux near the $\lambda_{\rm rest}=1212 - 1215\,$\AA\ region that the DLA alone underfits (compare top and bottom panels in Figure \ref{fig:app_igm}).     %
To reduce the dimensionality of the problem, in what follows we fix the proximity zone to the measured value (0.67 physical Mpc)\footnote{Note that a larger proximity zone would translate into a slightly larger $N_{\mathrm{HI}}$. If the proximity zone is forced to be $>0.8$ physical Mpc, all the fits to the data are worse than for smaller proximity zones.}.
Even then a combined model of IGM+DLA with two free parameters is degenerate.
To overcome this difficulty, we step through a number of IGM damping wing profiles using a fixed $\fhi$ (blue dashed lines in Figure \ref{fig:app_igm}). Then we perform a least-square regression to find the best DLA profile to match the data using as input the continuum already attenuated by the IGM. The result is that with a combined IGM+DLA model we always find a better fit to the absorption profile than using only a DLA.
Because neutral IGM neutral fractions $\fhi \gtrsim0.4$ have only been reported at $z\sim 7.5$ \citep{davies2018b, hoag2019} we find that unlikely to be the case at $z\sim 6.4$, which would also be in strong tension with constraints obtained toward other quasars and GRBs at comparable redshifts \citep[e.g., ][]{eilers2018,chornock2014,melandri2015}. We assume as our fiducial scenario an IGM that is 10\% neutral, in which case the best-fit DLA profile has a column density of $\log N_{\mathrm{HI}} = 20.68$. Nonetheless, we note that the cases where the IGM neutral fraction ranges from 5 to 50\% produce comparable good fits to the data (see Figure \ref{fig:app_igm}). We also note that $\fhi=0$ cannot be completely ruled out as the step in flux around $\lambda_{\rm rest}\sim 1214\,$\AA\ could be reproduced if one includes a saturated but otherwise weak (e.g.,  $\log N_{\mathrm{HI}}\sim 15.5$) $\lya$ absorption in that region. Indeed, such absorption features are frequently seen in DLAs and PDLAs at lower redshifts \citep[e.g.,][]{prochaska2005,prochaska2008}.
 Our fiducial value with its conservative uncertainty, $\log N_{\mathrm{HI}} = 20.68\pm 0.25$ (see Section \ref{sec:cont-blue}), encompasses the best-fit $N_{\mathrm{HI}}$ values found for all cases where $\fhi<0.38$ (see Figure \ref{fig:app_dlapars}).

\section{Testing for possible saturation}
\label{app:saturation}

 At the FIRE resolution ($\sim 50\,\kms$) it is possible that some of our lines suffer from some hidden saturation.   Here we will take a closer look at the absorption systems from Figure \ref{fig:metals}. The AODM can reveal potentially saturated or blended lines if the column densities for multiple lines of a single ion are significantly different. Only for \mgii\ and \siii\ do we have multiple transitions available.    The column densities of \mgii\ $\lambda$ 2803 and 2796 are consistent with each other but the column densities of \siii\ $\lambda$ 1526, 1304, and 1260 are significantly different, indicating potential issues.

\begin{figure}
\centering
\figurenum{12}
\plotone{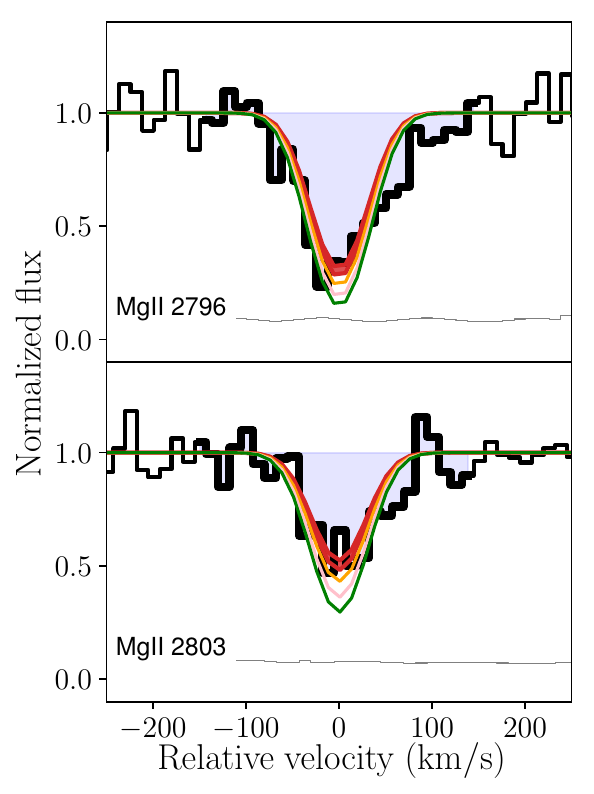}
\caption{
Velocity plot for \mgii\ $\lambda\lambda$2796,2803. The shaded red region shows the Voigt profile using their weighted mean column density derived using the AODM: $\log N_{\mathrm{Mg}}=13.37\pm 0.03$ (see Table \ref{tab:abundances}). The other lines show the expected absorption profiles for different column densities with increment of 0.1 dex. Both lines are consistent with each other; there are no evident signs of saturation for these transitions.
}
\label{fig:app_mg}
 \end{figure}

\begin{figure}
\centering
\figurenum{13}
\plotone{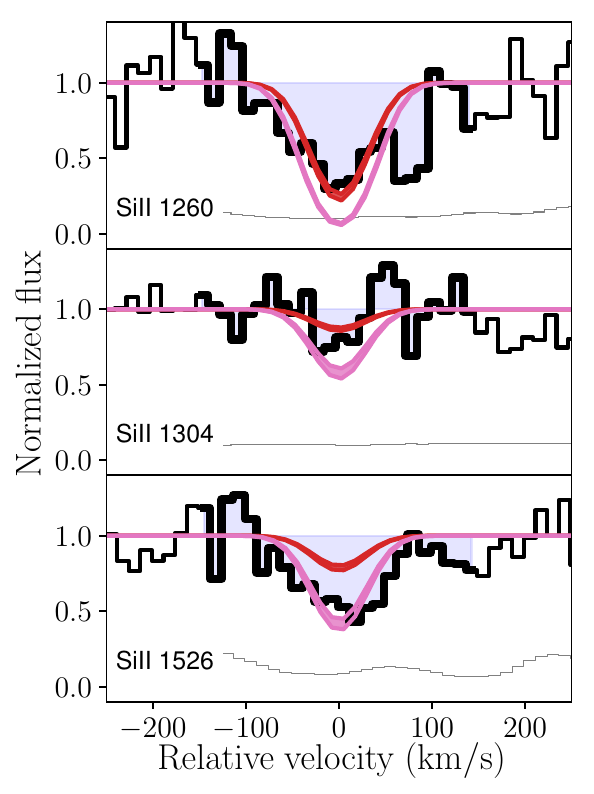}
\caption{
Velocity plot for \siii\ $\lambda$1260, 1304, and 1526. In all panels the shaded red region shows the Voigt profile using the column density derived using the AOD method for \siii\ 1260: $\log N_{\mathrm{SiII\,1260}}=13.53\pm 0.04$, while the shaded pink region shows the Voigt profile using the column density derived using the AODM for \siii\ 1526: $\log N_{\mathrm{SiII\,1526}}=14.15\pm 0.05$  (see Table \ref{tab:abundances}). The derived column densities are remarkably different, which is a sign of saturation or blended lines.  The column density of \siii\,1526 overpredicts the observed column density of both \siii\ 1260 and 1304. However, the \siii\ 1260 region is potentially contaminated by a \civ\ 1548 absorption line from a system at $z=5.0172$. Therefore, for the analysis of this work we treat the measured column density of \siii\ 1260 as an upper limit for Si (see Table \ref{tab:abundances}).
}
\label{fig:app_si}
 \end{figure}

 In Figure \ref{fig:app_mg} we show the velocity plot for \mgii\ $\lambda$2796, 2803 overlaid with a Voigt profile using the weighted mean column density derived using the AODM (see Table \ref{tab:abundances}). For all the Voigt profiles shown in this Appendix we use a referential Doppler parameter $b=25\,\kms$ but we note that the actual number does not have an important effect given that the profile is convolved with the $50\,\kms$ resolution of our data. As expected from the AODM analysis, the profile fits well both \mgii\ lines and therefore we treat the AODM measurements as robust.   As shown in Figure \ref{fig:app_si}, the case for the \siii\ lines is quite different. We overplot Voigt profiles using the column densities of the two strongest detections of \siii: $\log N_{\mathrm{SiII\,1260}}=13.53\pm 0.04$ in red and $\log N_{\mathrm{SiII\,1526}}=14.15\pm 0.05$ in pink. The \siii\ $\lambda$1526 clearly overpredicts the expected column densities for the other two \siii\ lines, including the weaker $\lambda1304$ transition, indicating that this line suffers from some unidentified contamination. Even though the \siii\ $\lambda$1260 column density seems consistent with the observed column density of \siii\ $\lambda$1304 at the S/N of our data, the region near \siii\ $\lambda$1260 is potentially contaminated by a \civ\ 1548 absorption line from a system at $z=5.0172$. Therefore, to be conservative, we consider the \siii\ $\lambda$1260 column density as an upper limit, and exclude \siii\ $\lambda$1526.

\begin{figure}
\centering
\figurenum{14}
\plotone{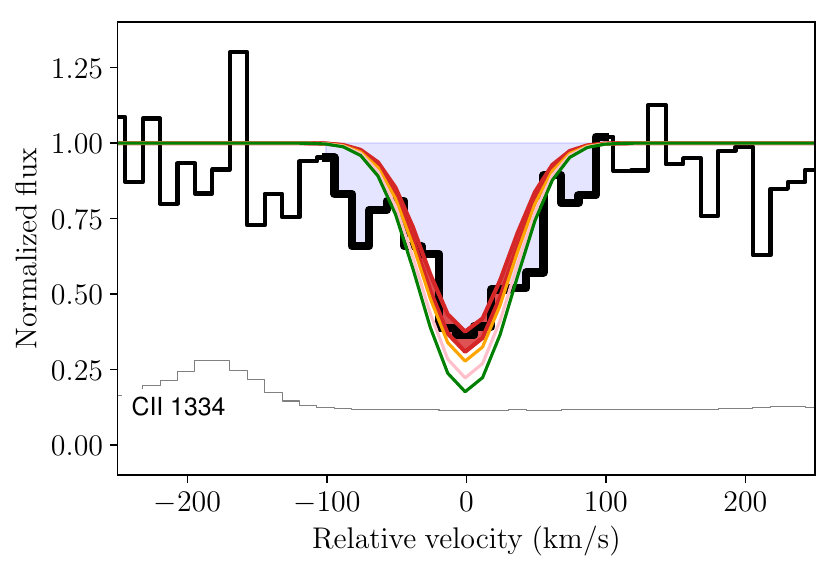}
\caption{
Velocity plot for \ciib\ $\lambda$1334. The shaded red region shows the Voigt profile using the column density derived using the AODM: $\log N_{\mathrm{CII}}=14.30\pm 0.05$ (see Table \ref{tab:ews}). The other lines show the expected absorption profiles for different column densities in increments of 0.1 dex. There is no sign of saturation for this transition.
}
\label{fig:app_cii}
 \end{figure}

\begin{figure}
\centering
\figurenum{15}
\plotone{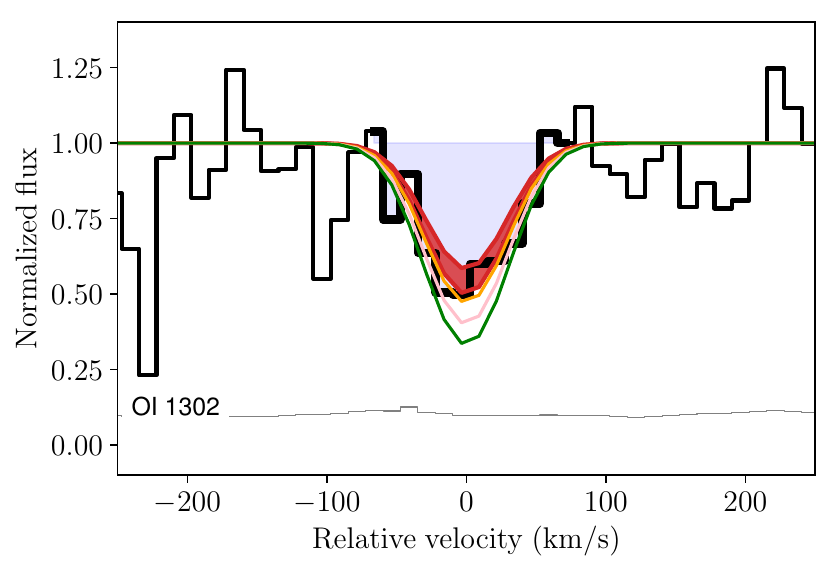}
\caption{
Velocity plot for \oi\ $\lambda$1302. The shaded red region shows the Voigt profile using the column density derived using the AODM: $\log N_{\mathrm{OI}}=14.45\pm 0.06$ (see Table \ref{tab:ews}). The other lines show the expected absorption profiles for different column densities in increments of 0.1 dex. There could be hidden saturation in this line sampled at this resolution (see orange line). Thus, to be conservative we increase the uncertainty of this measurement to 0.20 dex as that would still be consistent with the data.
}
\label{fig:app_oi}
 \end{figure}

\begin{figure}
\centering
\figurenum{16}
\plotone{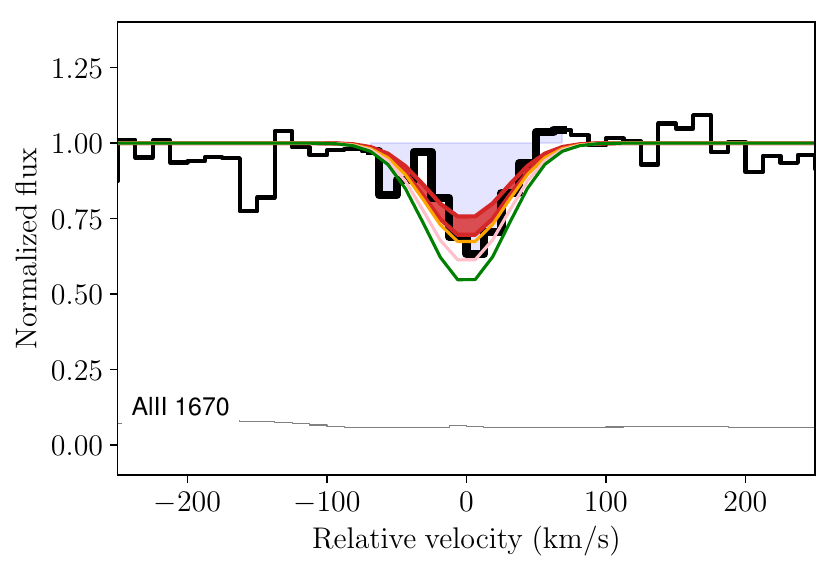}
\caption{
Velocity plot for \alii\ $\lambda$1670. The shaded red region shows the Voigt profile using the column density derived using the AODM: $\log N_{\mathrm{AlII}}=12.49\pm 0.06$ (see Table \ref{tab:ews}). The other lines show the expected absorption profiles for different column densities in increments of 0.1 dex. With our current data is not possible to rule out some of the higher column densities. Thus, to be conservative we increase the uncertainty of this measurement to 0.20 dex as that would still be consistent with the data.
}
\label{fig:app_al}
 \end{figure}

\begin{figure}
\centering
\figurenum{17}
\plotone{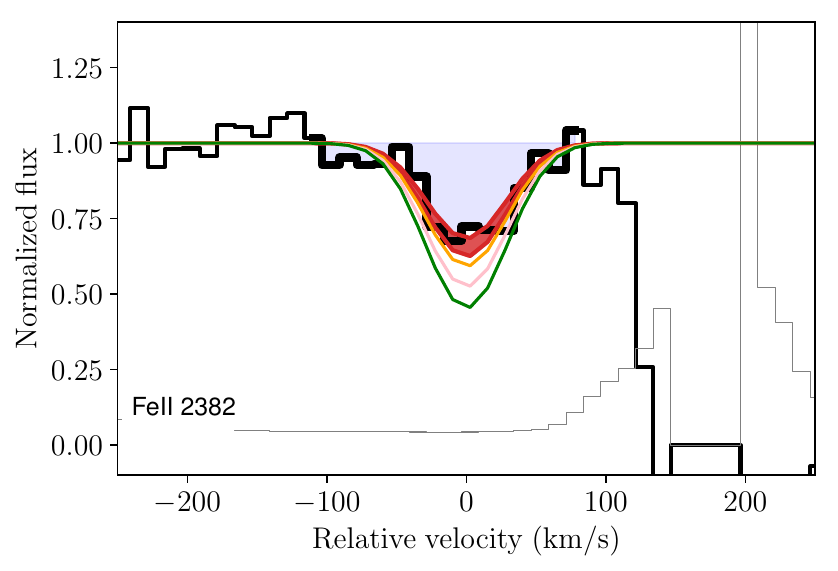}
\caption{
Velocity plot for \feii\ $\lambda$2382. The shaded red region shows the Voigt profile using the column density derived using the AODM: $\log N_{\mathrm{FeII}}=13.19\pm 0.05$ (see Table \ref{tab:ews}). The other lines show the expected absorption profiles for different column densities in increments of 0.1 dex. There is no sign of saturation for this transition.
}
\label{fig:app_fe}
 \end{figure}

 For ions where we only have a detection of a single line, the AODM does not provide information whether they could be contaminated or saturated. In Figures \ref{fig:app_cii}, \ref{fig:app_oi}, \ref{fig:app_al}, and \ref{fig:app_fe} we show the velocity plots and corresponding Voigt profiles using the AODM-derived column densities for \ciib\ $\lambda$1334, \oi\  $\lambda$1302, \alii\ $\lambda$1670, and \feii\ $\lambda$2382, respectively. We also overlay three additional Voigt profiles in each figure for different column densities with increment of 0.1, 0.2, and 0.3 dex from the AODM measurement as an attempt to identify potentially hidden saturation.  We find no evidence of saturation for \ciib\ $\lambda$1334 and \feii\ $\lambda$2382. On the other hand, with the resolution and sensitivity of our data cannot 100\% rule out the possibility that the column density of \oi\ $\lambda$1302 and \alii\ $\lambda$1670 could be slightly larger (see Figures \ref{fig:app_oi} and \ref{fig:app_al}). Therefore, to be conservative we have increased their column density uncertainties to 0.1 and 0.2 dex, respectively (see Table \ref{tab:ews}).

\end{document}